\documentclass[useAMS,usenatbib]{mn2e}

\usepackage{graphicx}
\usepackage{indentfirst}
\usepackage{amssymb}
\usepackage{widetext}

\renewcommand{\cos}{\mathrm{cos}}

\title[On the mean profiles of radio pulsars I: Theory of the propagation effects]
{On the mean profiles of radio pulsars I: Theory of the propagation effects}
\author[V. S. Beskin and A. A. Philippov]{V. S. Beskin$^{1,2}$\thanks{E-mail:
beskin@lpi.ru} and A. A. Philippov$^{2}$
\\
$^{1}$P.N.Lebedev Physical Institute, Leninsky prosp., 53, Moscow, 119991, Russia\\
$^{2}$Moscow Institute of Physics and Technology, Dolgoprudny, Institutsky per., 9,
Moscow region, 141700, Russia}

\begin{document}

\maketitle

\label{firstpage}

\begin{abstract}

We study the influence of the propagation effects on the mean profiles of radio pulsars using 
the method of the wave propagation in the inhomogeneous media describing by Kravtsov \& Orlov 
(1990). This approach allows us firstly to include into consideration the transition from 
geometrical optics to vacuum propagation, the cyclotron absorption, and the wave refraction 
simultaneously. In addition, non-dipole magnetic field configuration, drift motion of plasma 
particles, and their realistic energy distribution  are taken into account. It is 
confirmed that for ordinary pulsars (period \mbox{$P \sim 1$ s}, surface magnetic field 
\mbox{$B_0 \sim 10^{12}$ G}) and typical plasma generation near magnetic poles (the multiplicity 
parameter $\lambda  = n_{\rm e}/n_{\rm GJ} \sim 10^3$) the polarization is formed inside the light 
cylinder at the distance $r_{\rm esc} \sim 1000 R$ from the neutron star, the circular 
polarization being \mbox{$5$--$20$\%} which is just observed. The one-to-one correspondence 
between the signs of circular polarization and position angle ($p.a.$) derivative along the 
profile for both ordinary and extraordinary waves is predicted. Using numerical integration 
we now can model the mean profiles of radio pulsars. It is shown that the standard $S$-shape 
form of the $p.a.$ swing can be realized for small enough multiplicity $\lambda$ and large enough 
bulk Lorentz factor $\gamma$ only. It is also shown that the value of $p.a.$ maximum derivative, 
that is often used for determination the angle between magnetic dipole and rotation axis, 
depends on the plasma parameters {and could differ from the rotation vector model (RVM) 
prediction.} 

\end{abstract}


\section{Introduction}
\label{aba:sec1}
More than forty years after discovery, our understanding of pulsar phenomenon leaves an 
ambiguous impression. On the one hand, the key properties were understood almost immediately 
(see, e.g., the monographs by Manchester \& Taylor 1977; Lyne \& Graham-Smith 1998): the 
stable pattern of radio emission is related to the neutron star rotation, the energy source 
is the kinetic energy of its rotation, and the release mechanism has the electromagnetic
nature. The mean profiles of radio pulsars are well described by the hollow cone model 
(see below). Within four last decades enormous {amount} of observational data 
concerning polarization and other morphological properties of mean profiles was collected 
(Rankin 1983, 1990; Johnston et al. 2007; Weltevrede \& Johnston 2008; Hankins 
\& Rankin 2010; Keith et al. 2010; Yan et al. 2011). However, there is no agreement 
on the mechanism of coherent radio emission (Beskin 1999; Usov 2006; Lyubarsky 2008).

Any self-consistent theory of pulsar radio emission must include at least 
three main elements. First, it should describe a plasma instability that 
produces coherent radio emission. Second, the saturation of this instability 
that determines the intensity of the outgoing radio emission. Third, once the 
radiation is produced, its polarization properties are modified by the 
interaction with magnetospheric plasma and fields. These propagation effects 
should be accounted for to make a quantitative comparison of the theoretical 
predictions for the generation of radio emission with observational data.

There are several proposed mechanisms for the initial instability: an unstable flow of 
relativistic electron-positron plasma flowing along curved magnetic field lines 
(Goldreich \& Keeley 1971; Blandford 1975; Asseo et al. 1980; Beskin, Gurevich \& 
Istomin 1993, hereafter BGI); an instability caused by boundedness of the region of open 
field lines (Luo, Melrose \& Machabeli 1994; Asseo 1995); an instability connected with 
kinetic effects, that can be caused by non-equilibrium of the particles energy distribution 
function (mainly, anomalous Doppler effect in the region of cyclotron resonance); 
two-stream instability (Kazbegi, Machabeli \& Melikidze 1991); the instability 
connected with the nonstationarity of plasma particle production in the region of 
its generation (Lyubarskii 1996). 

The saturation mechanism, whose investigation requires involving the effects 
of a nonlinear wave interaction, is the most complex from the theoretical point 
of view. Therefore, it is not surprising that only a few researchers have managed 
to consider this question consistently (see, e.g., Istomin 1988). Finally, the 
processes of the wave propagation in pulsar magnetosphere
have not yet been investigated with sufficient detail either, although the part 
of theory that includes the propagation processes can be constructed using the 
standard linear methods of plasma physics.

{Recall that the present interpretation of the mean profiles of radio pulsars
is based on the ground of so-called hollow cone model (see, e.g., Manchester \& 
Taylor 1977). Within this approach the directivity pattern is assumed to repeat the
profile of the number density of secondary particles outflowing along the open field 
lines. As the secondary plasma cannot be generated near the magnetic pole (where the 
curvature photons radiating by primary beam propagate almost along magnetic field 
lines), the particle number density is to have the 'hole' in its space distribution
(Sturrock 1971; Ruderman \& Sutherland 1975).}

There are four assumptions in the hollow cone model: first, the emission is generated in 
the inner magnetospheric regions (where the magnetic field may be considered as a dipole); 
second, the emission propagates along the straight line; third, the cyclotron absorption 
may be neglected; and fourth, the polarization is determined at the emission point. Such 
basic characteristics of the received radio emission allow to determine the change of the 
position angle ($p.a.$) of the linear polarization along the mean profile (Radhakrishnan \& 
Cocke 1969)
\begin{equation}
p.a. = {\rm arctan} \left(\frac{\sin \alpha \sin \phi}
{\sin \alpha \cos \zeta \cos \phi - \sin \zeta \cos \alpha}\right).
\label{p.a.}
\end{equation}
Here $\alpha$ is the inclination angle of the magnetic dipole to the rotation axis, 
$\zeta$ is the angle between the rotation axis and the observer's direction, and $\phi$ 
is the pulse phase. 

As a result, the radiation beam width $W_{\rm r}$ itself and its statistical dependence on the 
period $P$ can be qualitatively explained under these assumptions (Rankin 1983, 1990). {As 
the pulsar radio emission is highly polarized (Lyne \& Graham-Smith 1998), one could check the 
validity of the relation (\ref{p.a.}) as well. As is well-known, in many cases the observed $p.a.$ 
swing is in good agreement with this theoretical prediction. Besides, the polarization observations 
show that the pulsar radio emission consists of two orthogonal modes, i.e., of two components which 
$p.a.$ differ by $90^{\circ}$ (Taylor \& Stinebring 1986). It is logically to connect two such 
components with two normal modes, ordinary and extraordinary ones, propagating in a magnetoactive 
plasma (Ginzburg 1961).} It is not surprising that the hollow cone model in its simplest realization 
is currently widely used for quantitative determination of the parameters of neutron stars.

At the same time, it is well known that, in general, three main assumptions are incorrect. 
First of all, after the paper by Barnard \& Arons (1986), it became clear that the ordinary 
wave {(i.e., the wave which electric field belongs to the plane containing the wave 
vector ${\bf k}$ and the external magnetic field ${\bf B}$)} does not propagate in a straight 
line, but deflects away from the magnetic axis. Subsequently, this effect was studied in detail 
by Petrova \& Lyubarskii (1998, 2000), it was an important element in the BGI theory. The correction 
to relation (\ref{p.a.}) connected with the aberration was determined by Blaskiewicz et al. (1991), 
but it was {rarely} used in analysis of the observational data as well. 

Further, the cyclotron absorption that must take place near the light cylinder (Mikhailovsky 
et al. 1982) turns out to be so large that it will not allow the radio emission to escape the 
pulsar's magnetosphere (see, e.g., Fussel et al. 2003). Finally, the limiting polarization effect 
had not been discussed seriously over many years, although it was qualitatively clear that this 
effect must be decisive for explaining of high degree of circular polarization, typically $5$--$20$\%. 

Indeed, in the region of the radio emission generation located at $10$--$100$ neutron star 
radii (these values result from the hollow cone model), the magnetic field is still strong 
enough so the polarization of the two orthogonal modes is indistinguishable from a linear
one. {For this reason, it is logical to conclude that the polarization characteristics 
cannot be formed precisely in the emission region, and the propagation effects are to play 
important role (cf. Mitra et al. 2009). Nevertheless}, in an overwhelming majority of the 
papers, Eqn. (\ref{p.a.}) is used to investigate the polarization.

Recall that the limiting polarization effect is related to the escape of radio emission 
from a region of dense plasma, where the propagation is well described in the geometrical 
optics approximation (in this case, the polarization ellipse is defined by the orientation 
of the external magnetic field in the picture plane), into the region of rarefied plasma, 
where the emission polarization becomes almost constant along the ray. This process was well 
studied (Zheleznyakov 1977; Kravtsov \& Orlov 1990) and was used successfully for numerous 
objects, for example, in connection with the problems of solar radio emission (Zheleznyakov 
1964). However, in the theory of pulsar radio emission, such problem has not been solved. 
Above the papers where the level $r = r_{\rm esc}$ at which the transition from the geometrical 
optics approximation to the vacuum occurs, was only estimated (see, e.g., Cheng \& Ruderman 1979; 
Barnard 1986), one can note only a few paper by Petrova \& Lyubarskii (2000) (these 
authors considered the problem in the infinite magnetic field), by Petrova (2001, 2003, 
2006), as well as the recent papers by Wang, Lai \& Han (2010, 2011).

The goal of our paper is to consider all three main effects (i.e., refraction, cyclotron 
absorption, and limiting polarization) simultaneously in a consistent manner for realistic 
case. Not only the plasma density but also the magnetic field decreases with increasing 
distance from the neutron star will be included into consideration. Also, the non-dipole 
magnetic field, the drift motion of plasma particles, and realistic distribution function 
of outgoing plasma will be taken into account. 

In section 2 both ordinary and extraordinary waves propagation in the pulsar magnetosphere is 
briefly considered. In addition, the hydrodynamic derivation of dielectric tensor of relativistic 
magnetized plasma is given. In section 3 the main parameters of our model are discussed. In 
section 4 we discuss the transition from geometrical optics to vacuum propagation, the cyclotron 
absorption, and the wave refraction. The results of numerical calculations of the mean profiles 
of radio pulsar are presented in section 5. Finally, in section 6 we discuss our main results.

It is necessary to stress that the main goal of this paper is in describing the
theoretical ground of propagation effects only. For this reason, we are not going to discuss
here in detail the theoretical predictions for real objects. This will be done in the separate
paper. On the other hand, the theory described below is independent on the emission mechanism 
and, hence, can be applied to any theory of the pulsar radio emission.

\section{Two orthogonal modes}
\label{aba:sec2} 

\subsection{On the number of outgoing waves}

\begin{figure}

\includegraphics[scale=0.4]{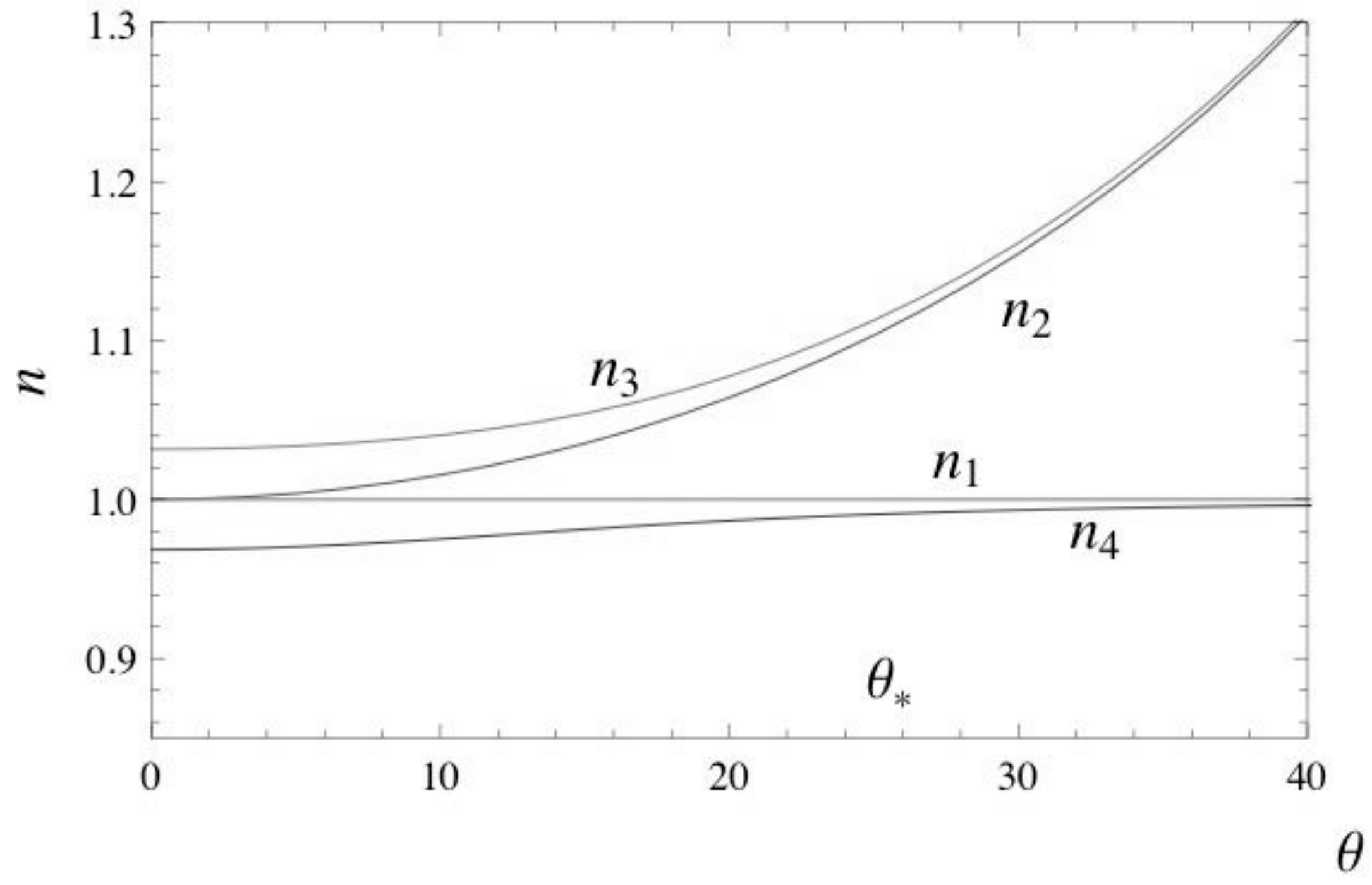}
\caption{Dependence of the refractive indexes $n$ on the angle $\theta$
between the wave vector ${\bf k}$ and external magnetic field ${\bf B}$
for $A_{\rm p} \gg 1$. The lower branch corresponds to the O-mode. 
The angle $\theta_{*}=\left<{\omega^2_{\rm p}}/{\omega^2\gamma^3}\right>^{1/4}$}
\label{fig00}
\end{figure}

As was already stressed, pulsar radio emission is highly polarized. The mean degree 
of linear polarization can reach $40$--$60$\%, and even 100\% in some subpulses (Lyne \&
Graham-Smith 1998). The analysis of the position angle demonstrates that in general the 
pulsar radio emission consists of two orthogonal modes, i.e., two modes in which their 
position angles differ by $90^{\circ}$. It is logical to connect them with the ordinary 
(O-mode) and extraordinary (X-mode) waves propagating in magnetized plasma (Ginzburg 1961). 

Starting from the pioneering work by Barnard \& Arons (1986),
three waves propagating outward in the pulsar magnetosphere
were commonly {considered} (see, e.g., Usov 2006; Lyubarsky 2008). 
{But in reality we have four waves propagating outwards.}
The point is that in the most of papers (see, e.g., Melrose \& Gedalin 1999) 
the wave properties were considered in the comoving reference frame in which 
the plasma waves propagating outward and backward are identical. But in the 
laboratory reference frame (in which the plasma moves with the velocity 
$v \approx c$) the latter wave is to propagate outward as well.

As shown on Fig.~\ref{fig00}, for zero angle $\theta$ between the wave vector ${\bf k}$ and 
external magnetic field ${\bf B}$ two of them, having refractive indices $n_1$ and $n_2$, 
correspond to transverse waves. For infinite external magnetic field $n_1 = n_2 = 1$. On the 
other hand, the waves $n_3$ and $n_4$ corresponds to plasma waves propagating in different
directions in the comoving reference frame.

Moreover, as was demonstrated by BGI, 
it is the fourth wave $n_4$ that is to be considered as the O-mode 
in the pulsar magnetosphere. It should be noted that it is valid for dense 
enough plasma in the radio generation domain for which $A_{\rm p} \gg 1$,
where
\begin{equation}
A_{\rm p}=\frac{\omega^2_{\rm p}}{\omega^2}<\gamma>.
\label{Ap}
\end{equation}
Here and below  $\omega_{\rm p} = (4 \pi e^2 n_{\rm e}/m_{\rm e})^{1/2}$ is the plasma 
frequency, $n_{\rm e}$ is the particle number density, $m_{\rm e}$ is the particle mass, 
and $\gamma$ is the particle Lorentz factor. Further, in what follows we assume that the 
particle distribution function $F_{e^{+}, e^{-}}(p)$ is one-dimensional. It results from 
the very high magnetic field in the vicinity of the neutron star where the synchrotron life 
time is negligible. In this case the brackets $<>$ denote both the averaging over the 
one-dimensional particle distribution function and the summation over the types of particles: 
\begin{equation}
<(...)> \, = \sum_{e^{+}e^{-}} \int (...) F_{e^{+},e^{-}}(p){\rm d} p.
\end{equation}

As a result, as shown in Fig. \ref{fig00}, for $A_{\rm p} \gg 1$ it is the wave $n_4$ that 
propagates as transverse O-mode at large angles $\theta \gg \theta_{*}$, i.e., at large 
distances from the neutron star. Here
\begin{equation}
\theta_{*}=\left<\frac{\omega^2_{\rm p}}{\omega^2\gamma^3}\right>^{1/4}.
\label{thetaQ}
\end{equation}
The second transverse wave for $\theta \gg \theta_{*}$ is again the X-mode $n_1$. 
Two other waves, $n_2$ and  $n_3$, for which the refractive index $n>1$, cannot 
escape from the magnetosphere as at large distances they propagate along the 
magnetic field lines (and due to Landau damping, see Barnard \& Arons 1986). 

In the hydrodynamical limit one can easily obtain the dispersion curves shown in Fig.~\ref{fig00} 
from the well-known dispersion equation in the limit of large magnetic field (see, e.g., 
Petrova \& Lyubarskii 2000) 
\begin{equation}
\left(1 - n^2 \cos^2\theta\right)
\left[1 - \frac{\omega_{\rm p}^2}{\omega^2 \gamma^3(1 - n v \cos\theta/c)^2}\right]
- n^2 \sin^2\theta = 0.
\end{equation}
For $\theta \ll \theta_{*}$ and for $\theta \gg \theta_{*}$ there are two transverse and 
two plasma waves, but for $A_{\rm p} \gg 1$ the nontrivial transformation from longitudinal 
to transverse wave takes place. This implies that in this case the mode $n_4$ can be emitted 
as a plasma wave, but it will escape from the magnetosphere as a transverse one.

As the refractive index $n_4$ differs from unity, the appropriate ordinary mode
deflects from the magnetic axis if $\theta  \leq \theta_{*}$. As was already mentioned, 
for the O-mode this effect takes place if $A_{\rm p} > 1$, i.e., for small enough 
distances from the neutron star $r < r_{\rm A}$, where
\begin{eqnarray}
r_{\rm A} \approx  10^{2} R \,  
\lambda_{4}^{1/3}  \,
\gamma_{100}^{1/3} \,
B_{12}^{1/3}  \,
\nu_{\rm GHz}^{-2/3} \,
P^{-1/3}.
\label{rA}
\end{eqnarray}
Here $R$, $P$, and $B_{12}$ are the neutron star radius, rotation period (in s), and 
magnetic field (in $10^{12}$ G), respectively. Accordingly,  $\gamma_{100} = \gamma/100$, 
$\nu_{\rm GHz}$ is the wave frequency in GHz, and $\lambda_{4} = \lambda/10^{4}$, where 
$\lambda = n_{\rm e}/n_{\rm GJ}$ is the multiplicity of the  particle creation near magnetic 
poles ($n_{\rm GJ} = \Omega B/2 \pi c e$ is the Goldreich-Julian number density). On the 
other hand, the transverse extraordinary wave with the refractive index \mbox{$n_1 \approx 1$} 
(X-mode) is to propagate freely. As the radius $r_{\rm A}$ is much smaller than the escape 
radius $r_{\rm esc}$ (Cheng \& Ruderman 1979; Andrianov \& Beskin 2010)
\begin{eqnarray}
r_{\rm esc} \approx  10^{3} R \,  
\lambda_{4}^{2/5} \,
\gamma_{100}^{-6/5} \,
B_{12}^{2/5} \,
\nu_{\rm GHz}^{-2/5} \,
P^{-1/5},
%\left(\frac{\theta}{0.2}\right)^{-6/5},
\label{resc}
\end{eqnarray}
one can consider the effects of refraction and limiting polarization separately.
In particular, this implies that one can consider the propagation of waves in the
region $r \sim r_{\rm esc}$ as rectilinear.

\subsection{Extraordinary wave}

\begin{figure}
\includegraphics[scale=0.45]{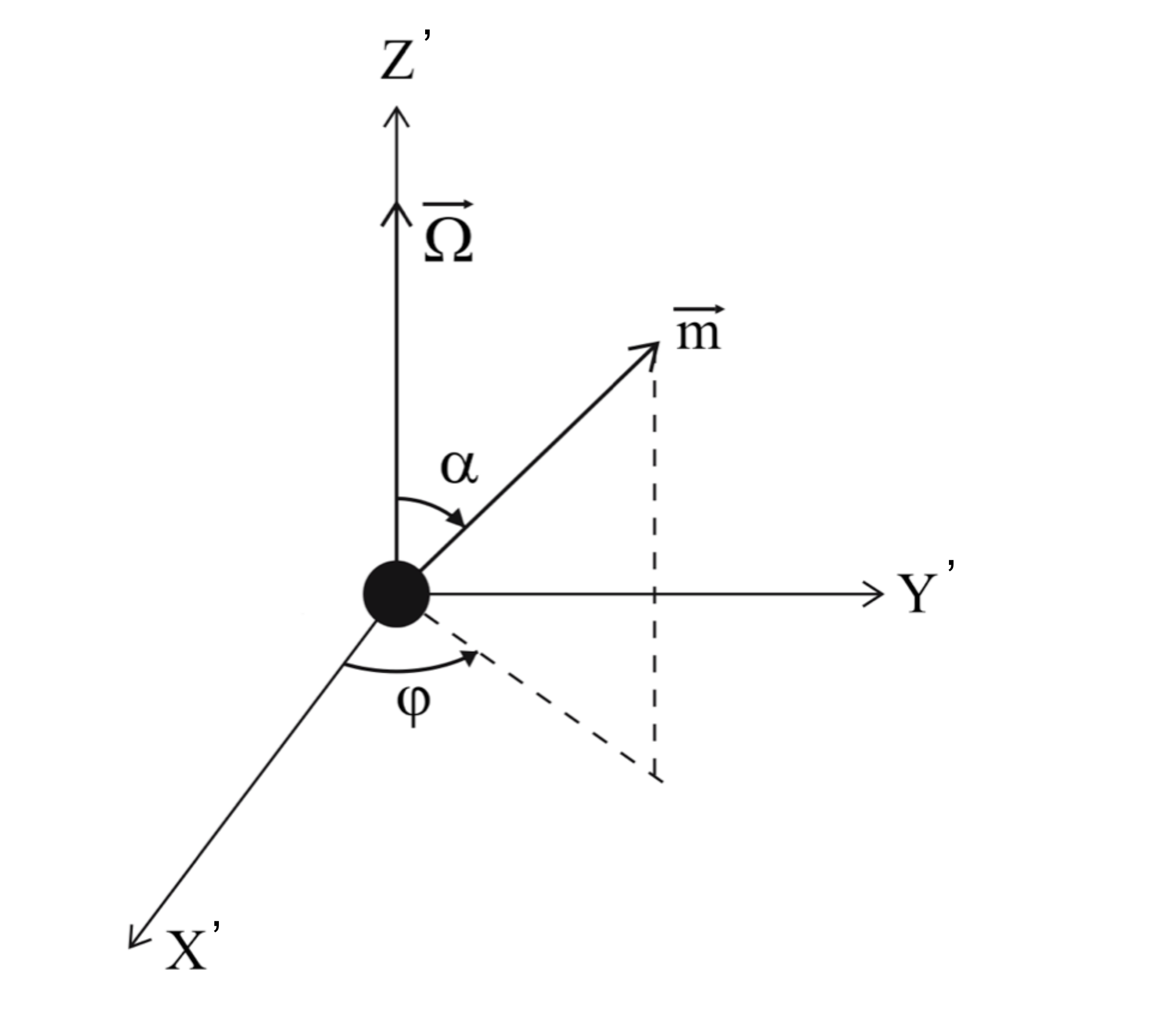}
\caption{$X'Y'Z'$ frame connecting with the neutron star rotation ($Z'$ axis)} 
\label{fig1s}
\end{figure}

\begin{figure}
\includegraphics[scale=0.45]{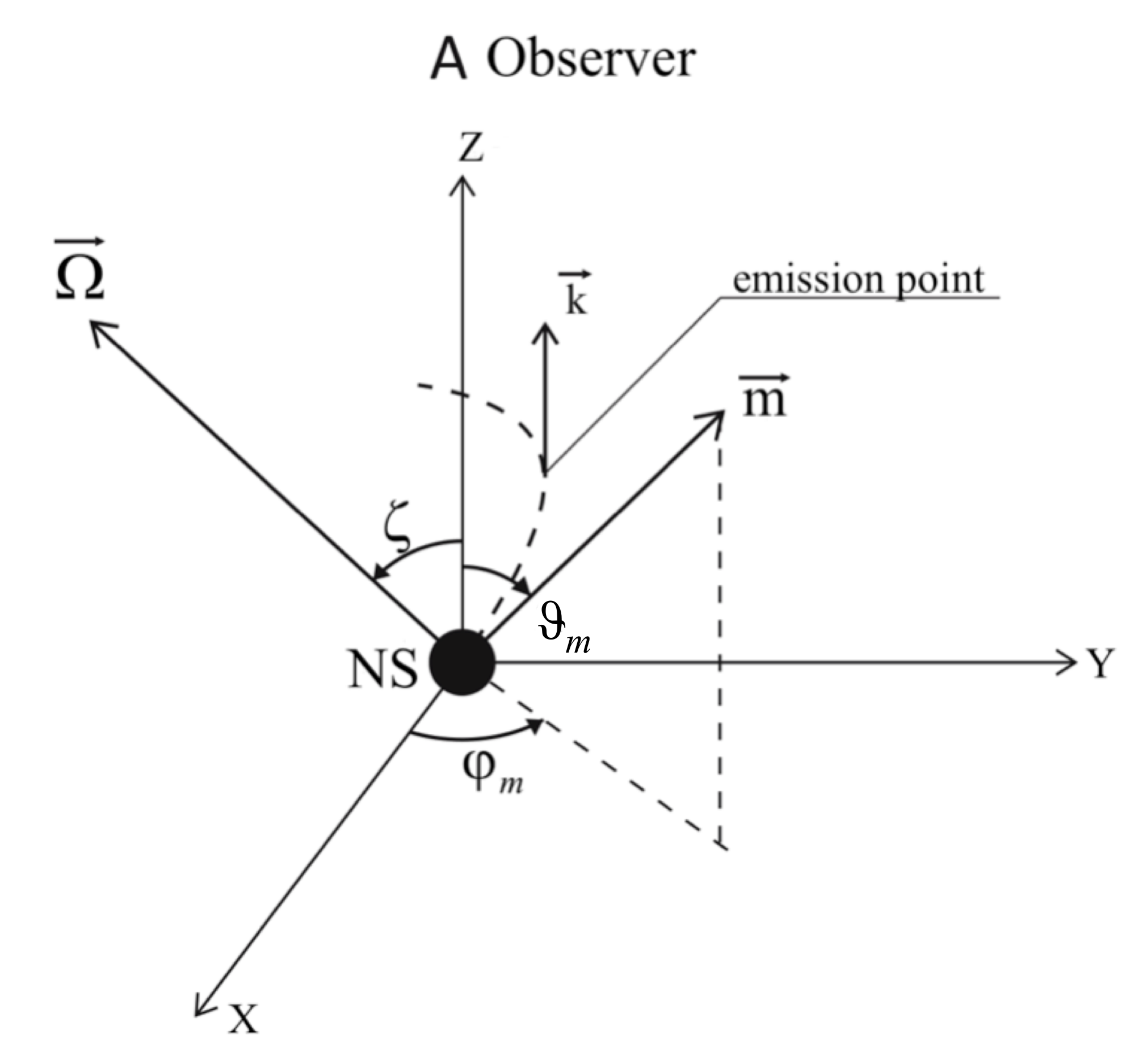}
\caption{$XYZ$ frame connecting with the line of sight ($Z$ axis).
The dashed curve indicates the magnetic field line} 
\label{fig2s}
\end{figure}

Below for simplicity we assume that both two outgoing modes are generated at the same 
heights $r_{\rm em}$ (few to tens NS radii), where the magnetic field can be 
considered as a rotating dipole
\begin{equation}
{\bf B}(\phi)=-\frac{{\bf m}(\phi)}{r^3}+\frac{3{\bf r}}{r^5}\left({{\bf m}(\phi),{\bf r}}\right).
\end{equation}
Here \(\phi = \Omega t\) is the corresponding pulsar rotation phase. 

{In what follows it is convenient to use two coordinate systems.
In a $X'Y'Z'$ frame ($Z'$-axis is along $\bf \Omega$; see Fig. \ref{fig1s})}, 
we have
\begin{equation}
{\bf m}(\phi)=\sin \alpha \cos \phi \, {\bf e}_{x'}+\sin \alpha \sin \phi \,{\bf e}_{y'}
+\cos \alpha \, {\bf e}_{z'}.
\end{equation}
{In a $XYZ$ frame ($Z$-axis is along the line of sight, $\bf \Omega$ lies in $XZ$ plane; see Fig. \ref{fig2s})} we have
\begin{eqnarray}
&&{\bf m}(\phi) = (\sin\alpha\cos\zeta\cos\phi-\sin\zeta\cos\alpha)  \, {\bf e}_{x}
+ \sin\alpha\sin\phi \, {\bf e}_{y} \nonumber \\ 
&&+(\cos\alpha\cos\zeta + \sin\alpha\sin\zeta\cos\phi){\bf e}_{z}.
\end{eqnarray}
{Therefore, in the $XYZ$ reference frame the spherical angles 
$\vartheta_{m}$ and $\phi_{m}$ of the vector \(\bf m\) are (see Fig. \ref{fig2s})}
\begin{eqnarray}
&&\cos \vartheta_{m} = \cos\alpha \cos\zeta+\sin\alpha \sin\zeta \cos\phi,\\
&&\tan \phi_{m} = - \frac{\sin \alpha \sin \phi}
{\sin \zeta \cos \alpha - \sin \alpha \cos \zeta \cos \phi}.
\end{eqnarray}
In the rotating vector model (RVM) the ${\it p.a.}$ is determined purely by the projection of 
magnetic field on the sky's plane, so it coincides with $\phi_{m}$. The sign of the arctan term 
is determined by the ${\it p.a.}$ measured {\it counter-clockwise} in the picture plane, as
is common in radio astronomy (Everett \& Weisberg 2001). 

As the aberration angle at the emission point is approximately $\Omega r_{\rm em}/{c}$, i.e., it is 
much smaller than the angular size of the emission cone $1/\gamma$, we can easily find the position 
of the emission point, at which the magnetic field line is along the line of sight. This 
point ${\bf r_{\rm em}} = (r_{\rm em},\theta_{\rm em},\phi_{\rm em})$ in the $XYZ$ frame is given by 
the spherical angles as
\begin{eqnarray}
&&\theta_{\rm em}=\frac{\vartheta_{m}}{2}-\frac{1}{2}\arcsin\left(\frac{1}{3}\sin\vartheta_{m}\right)
\approx \frac{\vartheta_{m}}{3},\\
&&\phi_{\rm em}=\phi_{m}.
\end{eqnarray}
Note that the impact angle \(\beta\) is the smallest angle between line of sight and magnetic moment 
${\bf m}$, is given by \(\beta=\alpha  - \zeta\). As a result, the trajectory of the extraordinary 
wave in the $XYZ$ frame is given by the simple relation
\begin{equation}
{\bf R} = {\bf r_{\rm em}}+r{\bf e}_{z}.
\end{equation}
This relation allows us to determine {the} magnetic field and all plasma characteristics along the ray.

\subsection{Ordinary wave}

Let us briefly review the main points of the theory of ordinary wave propagation 
(Barnard \& Arons 1986; Beskin et al. 1988; Petrova \& Lyubarsky 1990b). In the 
geometrical optics limit, the equations of motion of a ray are
\begin{equation}
\frac{{\rm d} {\rho}_{\perp}}{{\rm d} l} = \frac{\partial}{{\partial}k_{\perp}}\frac{k}{n_{j}},
\end{equation}
\begin{equation}
\frac{{\rm d} k_{\perp}}{{\rm d} l} = - \frac{\partial}{{\partial}\rho_{\perp}}\frac{k}{n_{j}},
\end{equation}
where ${\rho}_{\perp}$ is the distance from the magnetic dipole axis, $l \approx r$ is the 
coordinate along the ray, the index $\perp$ corresponds to the components perpendicular to 
the dipole axis, e.g., $\theta_{\perp} = k_{\perp}/k$, and $n_{j}$ are the corresponding 
refraction indices. Below in this subsection for simplicity the plasma density is assumed 
to be independent on the transverse coordinate $\rho_{\perp}$. 

As the refraction of the O-mode takes place at small distances from the neutron
star $r \ll r_{\rm A}$ (\ref{rA}), the expressions for refraction indices can be borrowed 
from the theory in the infinite magnetic field when the dielectric tensor of plasma 
has a form:
\begin{equation}
\nonumber \\
\varepsilon_{ij} =
\nonumber \\
\pmatrix{
1&& 0 && 0 \cr  
0&& 1 && 0 \cr
0&& 0 &&1-<\omega_p^2/(\tilde{\omega}^2\gamma^3)>\cr }.
\nonumber \\
\end{equation}
Here and below by definition
\begin{equation}
\tilde{\omega} = \omega - (\bf k, v).
 \end{equation}
{As the brackets $<>$ denote the averaging over the particle 
distribution function,  the singularities in the dielectric tensor coefficients in Cerenkov 
(and, below, in cyclotron) resonance vanish due to averaging over the wide particle 
energy distribution function}. 

As a result, for the ordinary mode the equations take the following form:
\begin{eqnarray}
\frac{{\rm d} \rho_{\perp}}{{\rm d} l} = \theta_{\perp} + \frac{\alpha_{B} 
- \theta_{\perp}}{2}\left(1-\frac{(\alpha_{B}-\theta_{\perp})^2}
{\left(\frac{16}{\omega^2}\left<\frac{\omega^2_{{\rm p}}}{\gamma^3}\right>
+(\alpha_{B} -\theta_{\perp})^4\right)^{1/2}}\right), \\
\frac{{\rm d} \theta_{\perp}}{{\rm d} l} = \frac{3}{4}\frac{\theta_{\perp}-\alpha_{B}}{l}
\left(1-\frac{(\alpha_{B}-\theta_{\perp})^2}{\left(\frac{16}{\omega^2}
\left<\frac{\omega^2_{{\rm p}}}{\gamma^3}\right>+(\alpha_{B}-\theta_{\perp})^4\right)^{1/2}}\right),
\end{eqnarray}
where $\alpha_{B}$ is the inclination angle of the magnetic field line to the magnetic axis. 
As was already mentioned, for large enough angles $\theta \gg \theta_{*}$ 
(\ref{thetaQ}) the ordinary wave propagates rectilinearly as well. From this condition 
and the solution of the equation above one can find
\begin{equation}
\theta_{\perp}(\infty) = \left(\frac{\Omega R}{c}\right)^{0.36}
\left(\frac{1}{\omega^2}\left<\frac{\omega^2_{{\rm p}0}}{\gamma^3}\right>\right)^{0.07}
f^{0.36}_{\rm em} \left(\frac{r_{\rm em}}{R}\right)^{0.15}.
\label{angle}
\end{equation}
Here $\omega_{{\rm p}0}$ is the plasma frequency near the star surface, 
and index 'em' corresponds to the quantities on the generation level. Besides, the 
dimensionless factor
\begin{equation}
f = \frac{c}{\Omega R}\left(\frac{l}{R}\right)^{-1}\sin^2\theta_{m} \sim 1,
\end{equation} 
where the angle $\theta_{m}$ is measured from the magnetic axis, determines the position of the 
radiation point within the polar cap. The angle $2 \theta_{\perp}(\infty)$ {then 
determines} the angular width of the {emission beam}. 
Finally, the "tearing off" level $l_{\rm t}$ defined 
by the condition $\theta = \theta_{*}$ is equals to
\begin{equation}
l_{\rm t} = 2R \left(\frac{\Omega R}{c}\right)^{-0.48}\left(\frac{1}{\omega^2}
\left<\frac{\omega^2_{p0}}{\gamma^3}\right>\right)^{0.24}f^{-0.48}_{\rm em} 
\left(\frac{r_{\rm em}}{R}\right)^{-0.20}.
\label{ltau}
\end{equation}
It gives
\begin{equation}
l_t \approx 40 R \, P^{0.24} \, \nu^{-0.48}_{\rm GHz} \,
\gamma^{-0.72}_ {100} \, B^{0.24}_{12} \, \lambda^{0.24}_{4}\, f^{-0.48}_{\rm em}
\left(\frac{r_{\rm em}}{R}\right)^{-0.2}.
\end{equation}
Thus, this level locates much {deeper} than the level of the formation of the outgoing 
polarization $r_{\rm esc} \sim 1000 \, R$ (\ref{resc}). 

\begin{figure}
\includegraphics[scale=0.21]{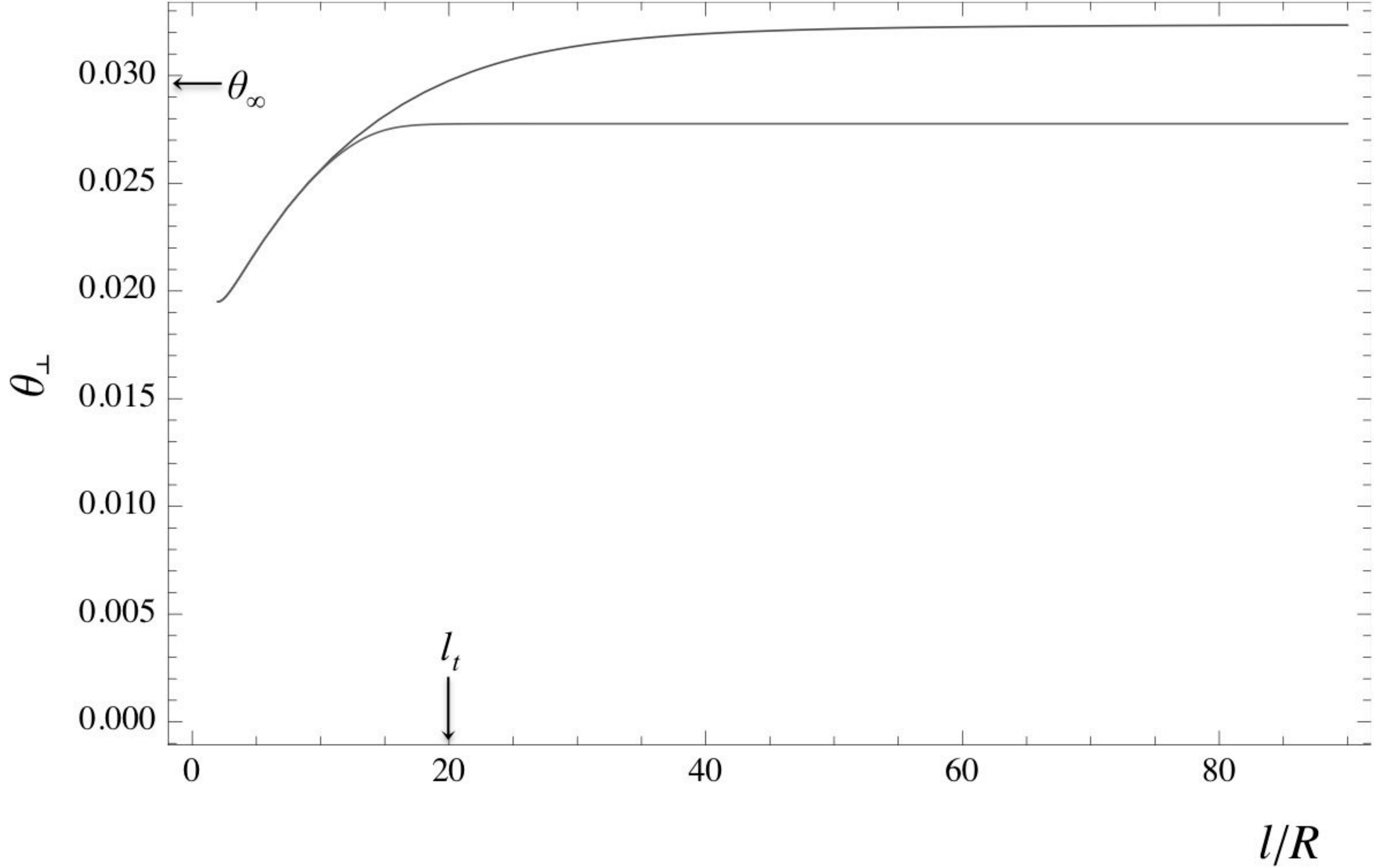}
\caption{Model simulation of the refraction equation for the absence (upper curve)
and the presence (lower curve) of the transverse number density gradient} 
\label{figrasp}
\end{figure}

{As was already stressed, these results were obtained 
for the case of neglecting transverse plasma density gradients. To check the validity of 
this approximation, we show on Fig.~\ref{figrasp} the characteristics of the wave propagation 
for two different density profiles. The upper curve corresponds to the absence of transverse 
gradients, and the lower one corresponds to the case of the hollow cone distribution that 
will be used everywhere below. As they are quite similar, the analytical results obtained 
above will be used. Also it is shown that the analytical estimates (\ref{angle}) and 
(\ref{ltau}) are correct enough.} In more detail the procedure we have used is described 
in Appendix A.

\subsection{Dielectric tensor}
\label{aba:sec3}

For reasonable parameters of the plasma filling the pulsar magnetosphere one can neglect 
the effect of curvature of magnetic field while considering the propagation of radio waves 
(see, e.g., Beskin 1999). On the other hand, as the level of the formation of the outgoing 
polarization $r_{\rm esc}$ (\ref{resc}) locates in the vicinity of the light cylinder 
(Cheng \& Ruderman 1979; Barnard 1986), it is necessary to include into consideration 
the nonzero external electric field (Petrova \& Lyubarskii 2000). In this paragraph $z$-axis 
is selected along the direction of magnetic field and the wave vector lies in $xz$-plane. 

Our goal is to find the permittivity tensor of relativistic plasma in perpendicular uniform 
magnetic and electrical fields. In the derivation we take into account the fact that in the
strong enough magnetic field the unpertubated motion of particles is the sum of the motion 
along the magnetic field lines and the electrical drift in the perpendicular direction:
\begin{equation}
{\bf V}_0=V_{\parallel}{\bf b}+{\bf U}.
\label{vdrift}
\end{equation} 
Here ${\bf b} = {\bf B}/B$ is the unit vector along the direction of magnetic field, and 
${\bf U}=c{[{\bf E,B}]}/{B^2}$ is the drift velocity. In what follows we will use another 
form of this equation
\begin{equation}
{\bf V}_0 = [{\bf \Omega}, {\bf r}] + c i_{\parallel}{\bf B}
\label{vdrift1}
\end{equation} 
resulting from the condition ${\bf E} + [{\bf V}, {\bf B}]/c = 0$ (BGI; Gruzinov 2006).
{Here $i_{\parallel}$ is the scalar function which will be determined below}.

To find the permittivity tensor $\varepsilon_{ij}$ we have to find the motion of plasma particles 
in the homogeneous fields pertubated by the plane wave. 
We start from linearized Euler equation:
\begin{equation}
\left(\frac{\partial}{\partial t}+{\bf V}_0\nabla\right)\delta{\bf P}
=e\left(\delta{\bf E}+\left[\frac{\delta{\bf V}}{c},{\bf B}\right]+\left[\frac{{\bf V}_0}{c},
\delta{\bf B}\right]\right),
\label{euler}
\end{equation}
\begin{equation}
\delta {\bf P} = 
{m_{\rm e}}\gamma\delta{\bf V}+{m_{\rm e}}{\gamma}^3\frac{({\bf V}_0,\delta V)}{c^2}{{\bf V}_0},
\end{equation}
and the relation between fields in the electromagnetic wave:
\begin{equation}
\delta{\bf B}=\frac{c}{\omega}[{\bf k},\delta {\bf E}].
\end{equation}
Writing now the clear relations for particle number density and electric current pertubations
\begin{equation}
\frac{\partial\delta {n_{\rm e}}}{\partial t}
+ {\rm div}({n_{\rm e}}\delta{\bf V}+\delta {n_{\rm e}}{{\bf V}_0})=0,
\label{continuity}
\end{equation}
\begin{equation}
\delta j_{i} = {n_{\rm e}e}\delta V_{i}+\delta {n_{\rm e}e} V_{0i} = \sigma_{ij}\delta E_{j},
\label{cond}
\end{equation}
where \(\sigma_{ij}\) is a conductivity tensor, one can determine the  permittivity tensor 
by the following relationship (Ginzburg, 1961)
\begin{equation}
\varepsilon_{ij} = \delta_{ij}+\frac{4\pi i}{\rm \omega}\sigma_{ij}.
\end{equation}
As a result, the expression for tensor  $\varepsilon_{ij}$ in the infinite magnetic field 
looks like (the full expressions can be found in Appendix B):

\begin{widetext}
\begin{equation}
\nonumber \\
\varepsilon_{ij} =
\nonumber \\
\pmatrix{
1 - <\frac{k_z^2U^2_{x}\omega^2_p\gamma^2_{U}}{\tilde{\omega}^2\gamma^3\omega^2}> && - <\frac{k_z^2U_{x}U_{y}\omega^2_p \gamma^2_{U}}{\tilde{\omega}^2\gamma^3\omega^2}> && - <\frac{k_z U_{x}\omega^2_p(\omega-k_xU_{x})\gamma^2_{U}}{\tilde{\omega}^2\gamma^3\omega^2}> \cr  
- <\frac{k_z^2U_{x}U_{y}\omega^2_p\gamma^2_{U}}{\tilde{\omega}^2\gamma^3\omega^2}>&& 1 - <\frac{k_z^2U^2_{y}\omega^2_p\gamma^2_{U}}{\tilde{\omega}^2\gamma^3\omega^2}> && - <\frac{k_z U_{y}\omega^2_p(\omega-k_xU_{x})\gamma^2_{U}}{\tilde{\omega}^2\gamma^3\omega^2}> \cr
- <\frac{k_z U_{x}\omega^2_p(\omega-k_xU_{x})\gamma^2_{U}}{\tilde{\omega}^2\gamma^3\omega^2}>&& - <\frac{k_z U_{y}\omega^2_p(\omega-k_xU_{x})\gamma^2_{U}}{\tilde{\omega}^2\gamma^3\omega^2}> &&1-<\frac{\omega_p^2(\omega-k_xU_{x})^2\gamma^2_{U}}{\tilde{\omega}^2\omega\gamma^3}>\cr }.
\nonumber \\
\end{equation}
\end{widetext}

Here now
\begin{equation}
\tilde{\omega}=\omega-k_xU_x-k_zv_{\parallel},
\end{equation}
and
\begin{equation}
\gamma_{U} = (1-U^2/c^2)^{-1/2}.
\end{equation}

\section{Magnetosphere model}

\subsection{Magnetic field structure}

As the formation of the outgoing polarization locates in the vicinity of the light cylinder, 
it is necessary to include into consideration the corrections to the dipole magnetic field 
which, actually, determines the disturbance of the $S$-shape form (\ref{p.a.}) of the 
${\it p.a.}$ swing. In this work we discuss the following models of magnetic field  
\begin{equation}
{\bf B} = {\bf B}_{\rm d} + {\bf B}_{\rm w},
\end{equation} 
where the field $\bf B_{\rm d}$ connects with the dipole magnetic field of the neutron star, 
and the
field $\bf B_{\rm w}$ corresponds to the outgoing wind.

For $\bf B_{\rm d}$ we discuss two possible models.
\begin{enumerate}
\item
The model of ''non-rotating dipole'', in which we neglect radiative corrections in the 
pre-exponential factors {(see Landau \& Lifshits 1975 for more detail)}:
\begin{eqnarray}
{B}_{r} & = & 2 \frac{|\bf{m}|}{r^3} \sin\theta \sin \alpha \, 
{\rm Re}\,[E_{\rm x}(r,\varphi, t)] + \\
&&2 \frac{|\bf{m}|}{r^3} \cos\theta \, \cos\alpha \nonumber,
\label{ll2a} \\
{B}_{\theta} & = & -\frac{|\bf{m}|}{r^3} \cos\theta \sin\alpha \, 
{\rm Re}\,[E_{\rm x}(r,\varphi, t)] + \\
&&\frac{|\bf{m}|}{r^3}  \sin\theta \, \cos\alpha \nonumber,
\label{ll2b} \\
{B}_{\varphi} & = & -\frac{|\bf{m}|}{r^3} \sin\alpha \, 
{\rm Re}\,[i \, E_{\rm x}(r,\varphi, t)],
\label{ll2c} 
\end{eqnarray}   
where here
\begin{equation}
E_{\rm x}(r,\varphi, t) = E_{\rm x}(r,\varphi -\Omega t) = \exp\left(i\frac{\Omega r}{c} +i\varphi -i\Omega t \right).
\end{equation}
As a result, since the ray propagates almost along the radius, i.e.,  $r\approx ct$, 
the full compensation of time and radial contributions in the factor 
$\exp\left(i\Omega r/c + i\varphi -i\Omega t \right)$ 
takes place, so the ray does not feel the dipole rotation.  
\item
The model of filled magnetosphere, which corresponds to a rigidly rotating dipole,
\begin{eqnarray}
{B}_{r} & = & 2 \frac{|\bf{m}|}{r^3} \sin\theta \sin\alpha \,
{\rm Re}\,[\exp\left(i\varphi -i\Omega t \right)] +\\
&&2 \frac{|\bf{m}|}{r^3} \cos\theta \, \cos\alpha,
\label{ll3a} 
\nonumber\\
{B}_{\theta} & = & -\frac{|\bf{m}|}{r^3} \cos\theta \sin\alpha \, 
{\rm Re}\,[\exp\left(i\varphi -i\Omega t \right)] + \\
&&\frac{|\bf{m}|}{r^3} \sin\theta \, \cos \alpha
\nonumber,
\label{ll3b} \\
{B}_{\varphi} & = & -\frac{|\bf{m}|}{r^3} \sin \alpha \, 
{\rm Re}\,[i \, \exp\left(i\varphi -i\Omega t \right)].
\label{ll3c} 
\end{eqnarray}
Such a magnetic field at the distances $r < R_{\rm L}$ was obtained by BGI (1993) 
and by Mestel et al. (1999) as a consistent solution of the force-free equation describing 
neutron star magnetosphere.  
\end{enumerate}  

As to the wind component, we use the following expressions corresponding to the so-called
''split monopole'' solution (Michel 1973; Bogovalov 1999)
\begin{eqnarray}
B_{r} & = &  \frac{\Psi_{\rm tot}}{2\pi r^2},
\label{Bb1} \\
B_{\varphi} & = &  -f_{\varphi}\frac{\Psi_{\rm tot}}{2 \pi R_{\rm L}} \, \frac{\sin \theta}{r}.
\label{Bb2}
\end{eqnarray} 
Here $\Psi_{\rm tot} = 2\pi f_{r} (\Omega/c) |\bf{m}|$  is the total magnetic flux through 
the polar cap, and $f_{r} \sim 1$ and $f_{\varphi} \sim 1$ are the dimensionless constants. 
We see that the former term {describes} the quasi-monopole radial magnetic field. 
Such a structure was obtained not only for the axisymmetric force-free (Contopoulos et al. 
1999; Timokhin 2006) and MHD (Komissarov 2006) numerical simulations but it describes well 
enough the magnetic field of the inclined rotator as well (Spitkovsky 2006). As we are 
actually {interested} in the disturbance of the dipole magnetic field inside the 
light cylinder only, we do not include here into consideration the switching of the radial 
field in the current sheet in the equatorial region. As the total magnetic flux through the 
polar cap depends only weakly on the inclination angle $\alpha$ (BGI; Spitkovsky 2006), we 
put here for simplicity $f_{r} = 1$ (for zero longitudinal current $f_{r}$ changes from 
1.592 to 1.93).

Besides, the latter term ${B}_{\varphi}$ (\ref{Bb2}) corresponds to the toroidal 
magnetic field connected with the longitudinal electric current flowing in the 
magnetosphere. It is well-known that to support the MHD (in particular, force-free) 
outflow up to infinity the the total longitudinal current $I$ is to be close to the 
Michel (1973) current $I_{\rm M} = \Omega \Psi_{\rm tot}/4\pi$ (Contopoulos et al. 
1999). It corresponds to $f_{\varphi} \approx 1$. On the other hand, {for realizing} this 
current for inclined rotator with dipole magnetic field, it is necessary to suppose 
that the current density $j_{\parallel}$ is much larger than the local 
Goldreich-Julian current $j_{\rm GJ} \approx \Omega B \cos \alpha/2\pi$ (Beskin 2010). 
As it is not clear whether the Michel current $I_{\rm M} > I_{\rm GJ}$ can be realized 
in the pulsar magnetosphere, in what follows the parameter $f_{\varphi}$ can be 
considered as a free one. 

\begin{figure}
\includegraphics[scale=0.4]{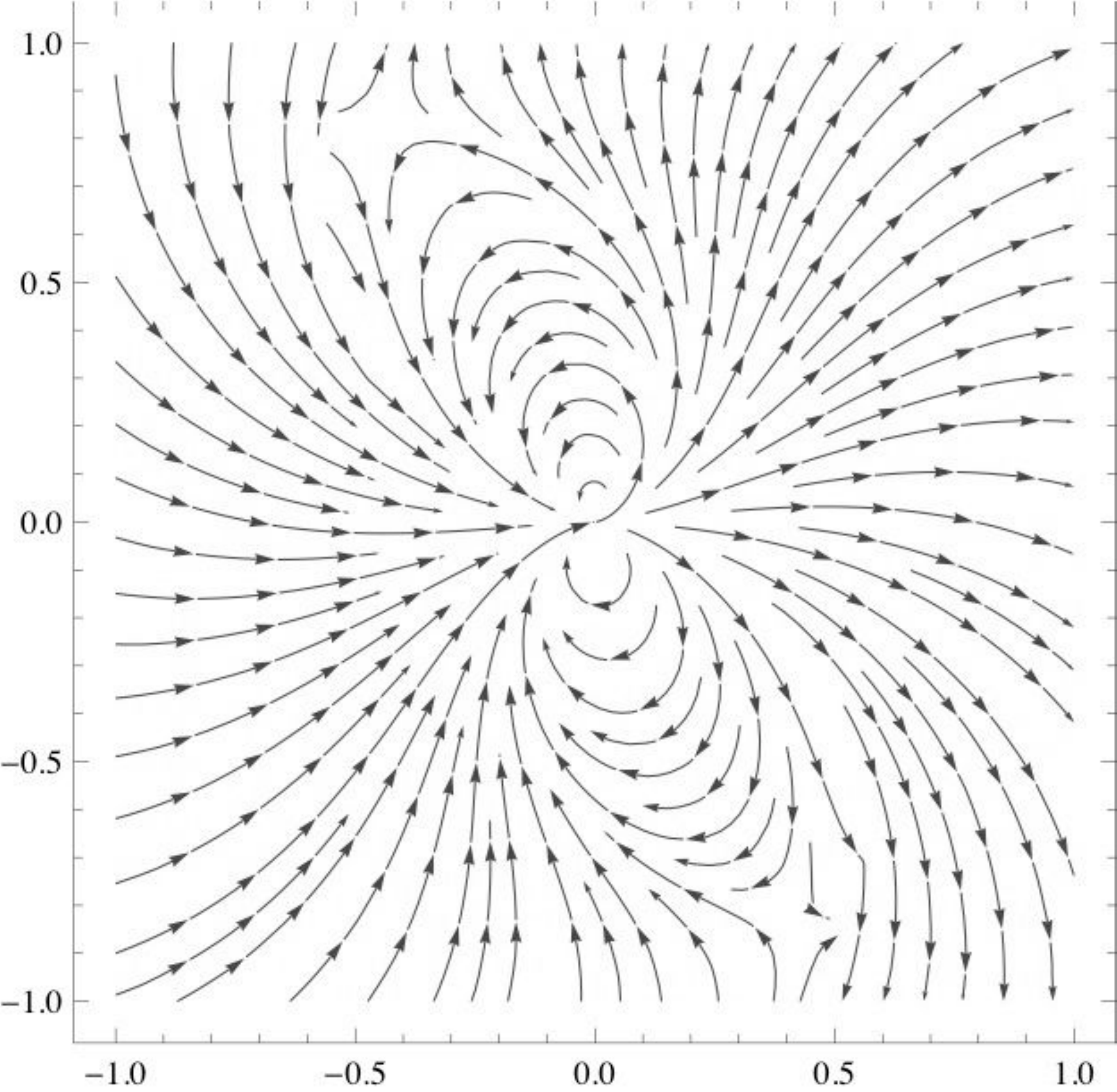}
\caption{Structure of the magnetic field lines in equatorial plane in corotating frame for the 
orthogonal rotator $\chi = 90^{\circ}$. The magnetic field is considered as the sum of dipole 
and ''split-monopole'' (model C). The scale corresponds to the light cylinder 
$R_{\rm L} = c/\Omega$} 
\label{figmagnfield}
\end{figure}

In Table~\ref{table00} we present the notation of the models which will be used in what follows.
Magnetic field structure for model C (and for orthogonal rotator) is shown in 
Fig~\ref{figmagnfield}. It {is qualitatively similar to} the numerical model obtained 
by Spitkovsky (2006).

\begin{table}
\caption{Models of the magnetic field structure}  
\centering
\begin{tabular}{|l|c|c|c|}
  \hline
model    & A                  &  B                & C                \\
  \hline
dipole   & $i$               & $ii$             & $ii$              \\
  \hline
wind     & $f_{\varphi} = 0$  & $f_{\varphi} = 0$ & $f_{\varphi} = 1$ \\
  \hline
\end{tabular}
\label{table00} 
\end{table}

\subsection{Plasma number density}

\begin{figure}
\includegraphics[scale=0.28]{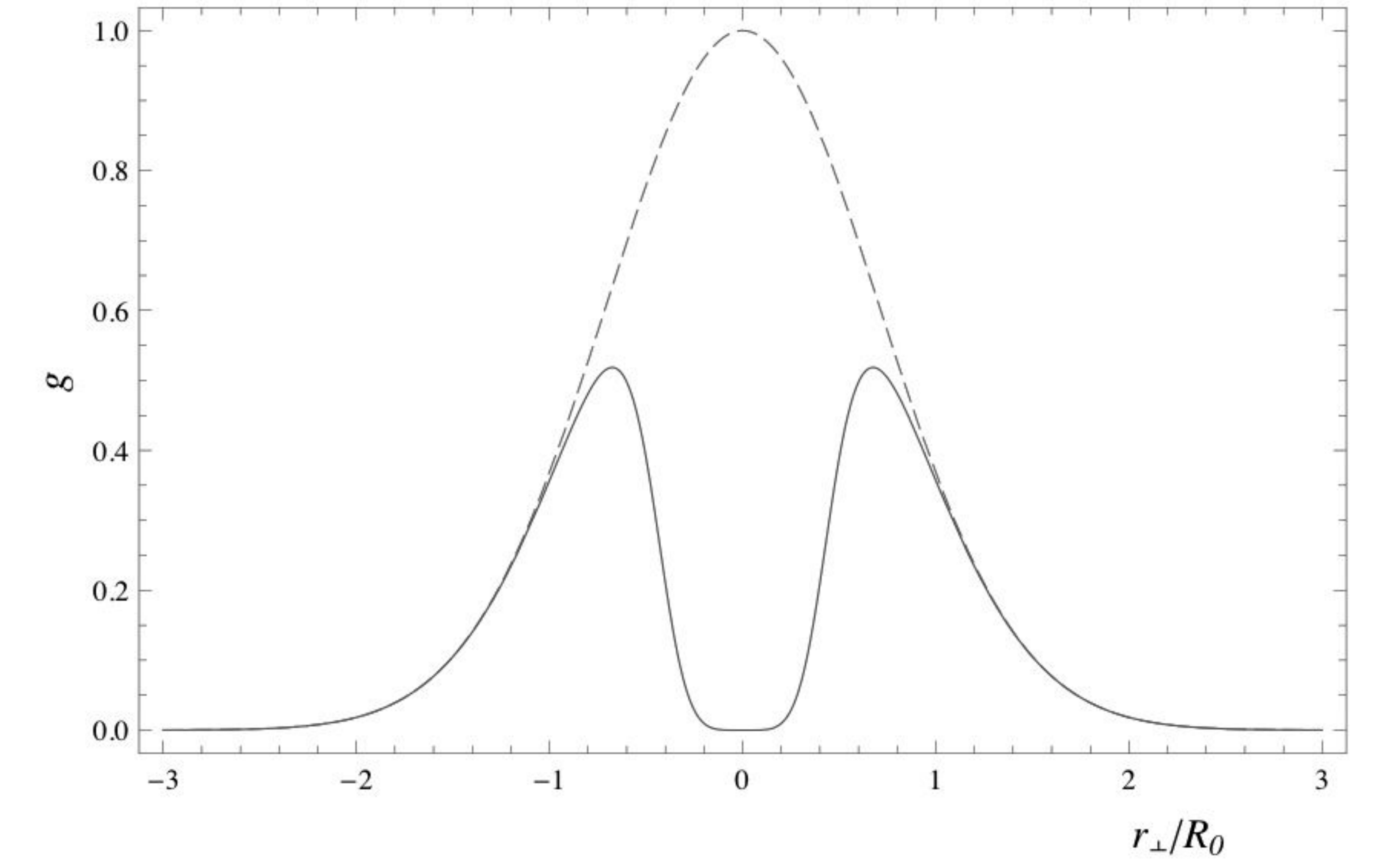}
\caption{Plasma space distribution function $g(r_{\perp})$ on the polar cap as a function of the
distance $r_{\perp}$ from the magnetic axis for \mbox{$f_0$ = 0.25.} The dashed line corresponds to 
$f_0 = 0$} 
\label{fig0}
\end{figure}

Recall the well-known property of the one-photon particle production in a strong magnetic 
field: the secondary particles are produced only if the photon moves at large enough angle 
to the magnetic field line (Sturrock 1971; Ruderman \& Sutherlend 1975; Arons \& Scharlemann 
1979). Since the relativistic particles near the neutron star surface can move only along the 
field lines with a Lorentz factor $\gamma=(1-v_{\parallel}^2/c^2)^{-1/2}$ ($v_{\parallel}$ 
is the particle velocity along the magnetic field), the hard gamma-quanta emitted through 
curvature mechanism also begin to move along the field lines. As a result, the production 
of secondary particles will be supressed near the magnetic poles, where the magnetic field 
is nearly rectilinear. Therefore, one would expect the secondary plasma density to be 
suppressed in the central region of the open field lines (see Fig. \ref{fig0}). It is this 
property that  lies in the ground of the hollow cone model.

Bellow we assume that the plasma number density on a polar cap is known. It is convenient to 
rewrite it in the form
\begin{equation}
n_{\rm e}(\theta_{m}, \varphi_{m}) = \lambda g(\theta_{m}, \varphi_{m}) 
n_{\rm GJ}^{(0)}.
\end{equation}
Here $n_{\rm GJ}^{(0)} = \Omega B/2\pi c e$ is the amplitude of the Goldreich-Julian number 
density, i.e., it does not depend on the inclination angle $\alpha$. Further, the multiplicity 
parameter \mbox{$\lambda = n_{\rm e}/n_{\rm GJ}^{(0)}$} determining the efficiency of the 
pair creation is (Daugherty \& Harding 1982; Gurevich \& Istomin 1985; Istomin \& Sobyanin 
2009; Medin \& Lai 2010)
\begin{equation}
\lambda \sim 10^{3} -10^{4}.
\end{equation}
Finally, the dimensionless factor $g(\theta_{m}, \varphi_{m}) \sim 1$ describes the real number 
density of the secondary plasma in the vicinity of the neutron star surface as a function of 
magnetic pole angles $\theta_{m}$ and $\varphi_{m}$.

The procedure described below allows us to determine the properties of the outgoing radiation
for arbitrary number density $n_{\rm e}$ within the polar cap. For illustration we consider
an axially symmetric distribution
\begin{equation}
g(f) = \frac{f^{2.5}\exp(-f^2)}{f^{2.5}+f_0^{2.5}},
\end{equation}
where $f = r_{\perp}^2/R_0^2$ is the dimensionless distance to the magnetic axis, 
$R_0 = (\Omega R/c)^{1/2} R$ is the polar cap radius, and the parameter $f_0$ 
describes the hole size in space plasma distribution (see Fig.~\ref{fig0}). 

As was already stressed, in this paper we are going to discuss the theoretical
ground of propagation effects only. For this reason we assume here for simplicity
the intensity of radio emission in the emission region to be proportional to the 
number density of outgoing plasma. In reality the directivity pattern may differ 
drastically from the particle profile.

To determine the number density $n_{\rm e}$ of the outgoing plasma in the arbitrary point of the 
magnetosphere, we use quasi-stationary formalism, which is valid for quantities that are 
functions of \(\phi-\Omega t\). For such functions all time derivatives can be 
reduced to spatial derivatives by the following rules (Beskin 2009):
\begin{equation}
\frac{\partial}{\partial t}Q = -\Omega\frac{\partial}{\partial \phi}Q,
\end{equation}
\begin{equation}
\frac{1}{c}\frac{\partial}{\partial t}{\bf V}=\nabla \times[{\bf{{\beta}_R}},{\bf V}]
-(\nabla\bf V){\bf{{\beta}_R}},
\end{equation}
for any scalar ($Q$) and vector (${\bf V}$) functions. Here
\begin{equation}
{\bf{{\beta}_R}}=\frac{\bf[\Omega,\bf\rm r]}{c}.
\end{equation}
Using now the continuity equation
\begin{equation}
\frac{\partial n_{\rm e}}{\partial t}+{\rm div}(n_{\rm e} {{\bf V}_0})=0,
\end{equation}
where the velocity \({\bf V}_0\) is given by Eqn. (\ref{vdrift1}), one can obtain
\begin{equation}
({\bf B}\nabla)(n_{\rm e} i_{\parallel})=0.
\label{Bn}
\end{equation}
Hence, the  product 
\begin{equation}
n_{\rm e} i_{\parallel} = \rm{const}
\end{equation} 
remains constant along the field lines (BGI, Gruzinov 2006).
Taking now into account only the first order by $\Omega r/c$ and assuming that the velocity 
of the outflowing particles is close to the light velocity $c$, we finally 
obtain
\begin{equation}
i_{\parallel}=\frac{1}{B}[1-(\bf b,{\bf{{\beta}_R}})].
\end{equation}

\begin{figure}
\includegraphics[scale=0.13]{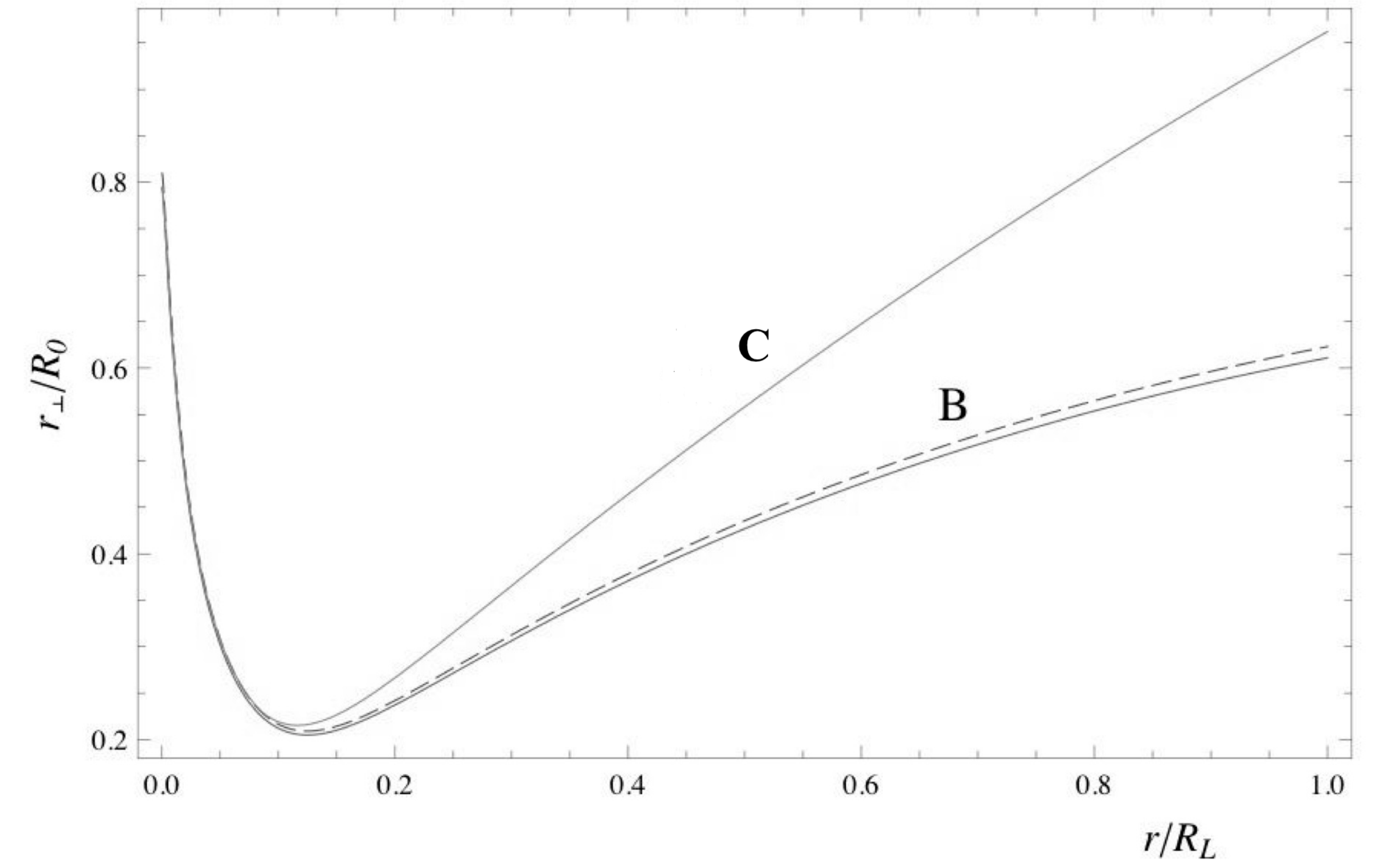}
\caption{Dependences of the distance of the foot points to the magnetic axis within the
polar cap as a function of the distance $r$ along the ray  for rotating dipole 
(model B) and for magnetic field including the monopole wind (model C) for the phase point 
$\phi =-5^{\circ}$. Dashed curve corresponds to analytical expression (\ref{analyte})} 
\label{figres0}
\end{figure}

Thus, to determine the number density $n_{\rm e}$ at an arbitrary point along the ray
trajectory it is enough to know the number density and magnetic field $B$ at the base
of a given field line on the neutron star surface. But for this it is necessary 
to produce the back integration along the field line from any point along the
trajectory up to the star surface. 

In Fig.~\ref{figres0} we show the dependences of the distance of the 
foot points to the magnetic axis $r_{\perp}$ within the polar cap as a function of the 
distance $r$ along the ray for rotating dipole (model B) and for magnetic 
field including the monopole wind (model C) for the phase point $\phi =-5^{\circ}$. 
As for model B the appropriate value can be obtained analytically (dashed curve)
\begin{equation}
r_{\perp} = R \sqrt{\frac{R}{r}} \, \sin \psi_m,
\label{analyte}
\end{equation}
where $\psi_m$ is the angle between local point on the ray vector and momentary magnetic axis, 
one can conclude that the precision of our procedure is high enough.

\subsection{Energy distribution and cyclotron absorption}

\begin{figure}
\includegraphics[scale=0.28]{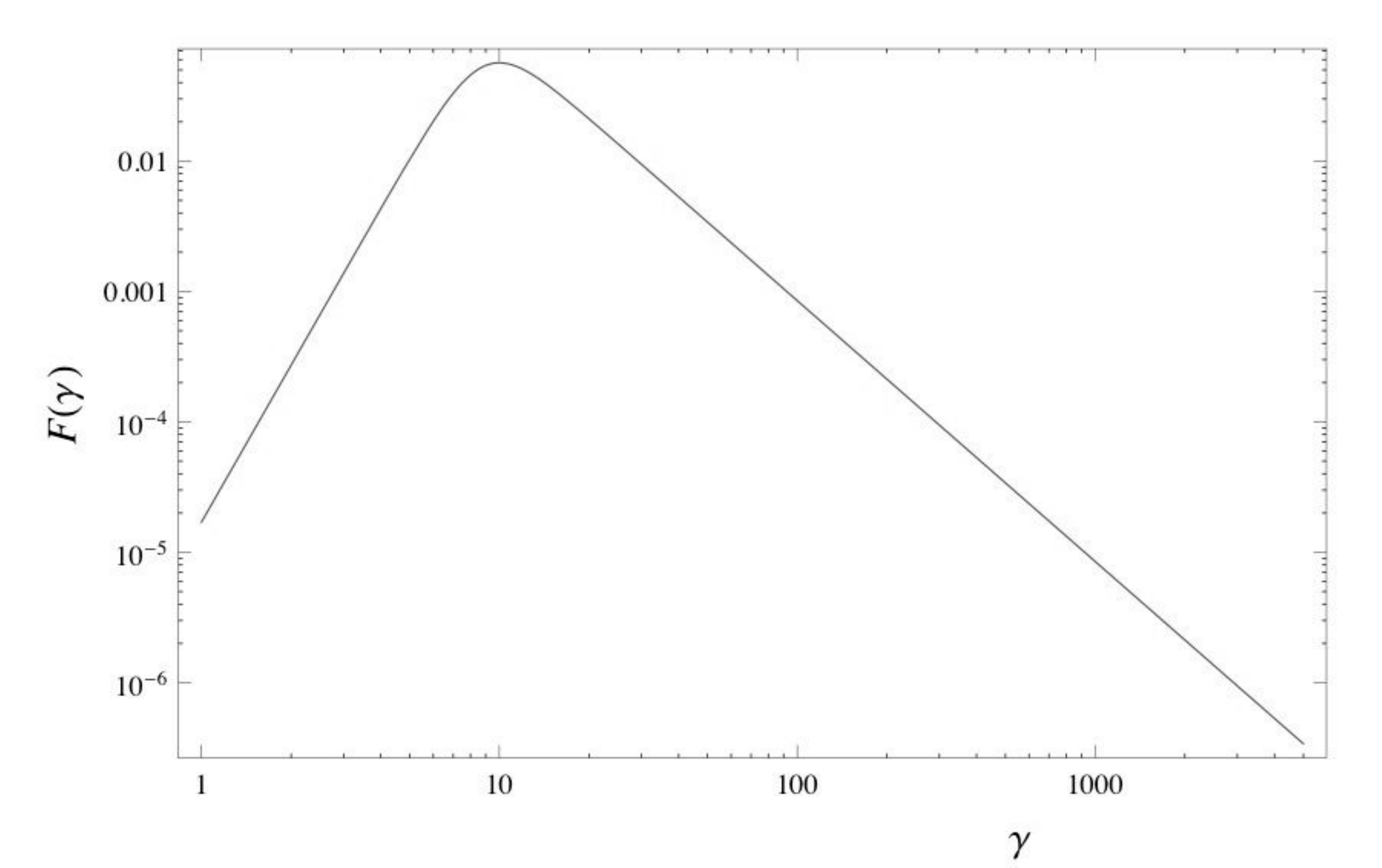}
\caption{Particle energy distribution function} 
\label{fig1}
\end{figure}

As was shown by Andrianov \& Beskin (2010), for large enough shear of the magnetic
field along the ray (as will be shown below, this condition does hold in the pulsar
magnetosphere), all the polarization characteristics of outgoing radiation depend 
on the diagonal components of dielectric tensor, that are not sensitive to the 
difference of the $e^{+}e^{-}$ distribution functions. This fundamental property
allows us to consider the electron and positron energy distribution functions to be 
identical. Our particular {choice} is (see Fig.~\ref{fig1})
\begin{equation}
F(\gamma) = \frac{6\gamma_0}{2^{1/6}\pi} \, \frac{\gamma^4}{2 \gamma^6 + \gamma^6_0}.
\label{distrfunc}
\end{equation}
This distribution has the maximum for $\gamma = \gamma_0$ that is assumed as a typical 
Lorentz-factor of plasma particles, and has power-law spectrum $\gamma^{-2}$ for large 
Lorentz-factors $\gamma \gg \gamma_0$. Thus, it models well enough the energy distribution 
function obtained numerically (Daugherty \& Harding 1982; BGI). 

\begin{figure}
\includegraphics[scale=0.3]{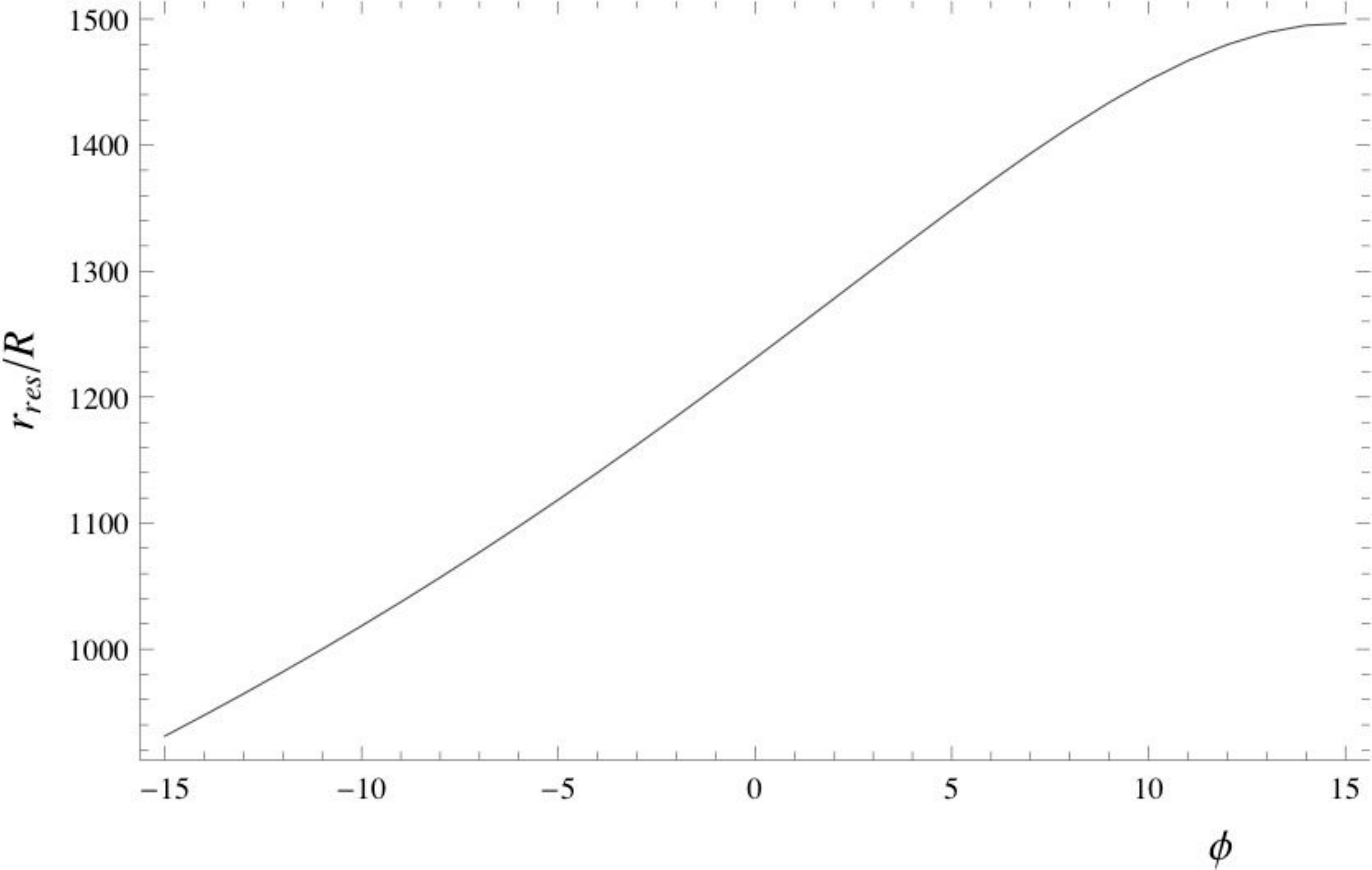}
\caption{Cyclotron resonance radius as a function of the pulsar rotation phase for  
$\nu = \rm{1GHz}$, $\gamma_{0} = \rm{100}$, and magnetic field model B} 
\label{figres}
\end{figure}

Finally, consider the cyclotron absorption taking place in the region where the
condition $\omega_{B} = \gamma \gamma_{U}\tilde{\omega}$ holds (see Appendix B
for more detail). As is well-known, the cyclotron resonance locates at the distances
\begin{equation}
r_{\rm res} \approx 2 \, 10^3 R \, 
\nu^{-1/3}_{\rm GHz}\gamma^{-1/3}_{100} B^{1/3}_{12}\theta^{-2/3}_{0.1}
\label{cycres} 
\end{equation}
comparable with the escape radius $r_{\rm esc}$ (\ref{resc}) (Mikhailovsky et al. 
1982). It implies that these two effects are to be considered simultaneously. On 
the other hand, as was already stressed, in this region one can neglect the wave 
refraction, i.e., to put ${\rm Re}\,[n] = 1$. {Remember that the estimate 
(\ref{cycres}) was obtained for zero drift velocity $U = 0$. Nevertheless, as one 
can see on Fig.~\ref{figres}, the result of calculation for real case is in qualitative 
agreement with estimation ({\ref{cycres}}){\footnote{Here the delta-function for particle 
energy distribution was assumed, because in the case of distribution function (\ref{distrfunc}) 
there is the wide zone of cyclotron resonance.}}.}

As a result, the intensity of outgoing radiation can be determined as
\begin{equation}
I_{\infty} = I_0 \exp(-\tau),
\label{intens}
\end{equation}
where $I_{0}$ is the intensity in the emission region, and the optical depth 
$\tau = 2 \omega/c \int {\rm Im}\, [n] \, {\rm d}l$ can be found using the clear
relation
\begin{equation}
{\rm Im}\,[n] \approx {\rm Im}\,[\varepsilon_{y'y'}]/2.
\end{equation}
As a result, we have
\begin{eqnarray}
\tau & \approx  & \frac{4\pi^2e^2}{m_{\rm e}c}\int\limits_0^\infty\int\limits_0^\infty 
n_{\rm e}(l) \frac{\tilde{\omega}}{\omega} F (\gamma) 
\delta\left(|\omega_B|\sqrt{1 - \frac{U^2}{c^2}}-\gamma\tilde{\omega}\right) 
{\rm d}\gamma {\rm d}l
\nonumber\\
&=&\frac{4\pi^2e^2}{m_{\rm e}c}\int\limits_0^\infty n_{\rm e}(l)\frac{1}{\omega} 
F \left(\frac{|\omega_B|\sqrt{1 - U^2/c^2}}{\tilde{\omega}}\right) {\rm d}l.
\end{eqnarray}
Here we use the approximation \(v^2_{\parallel}/c^2 \approx 1 - U^2/c^2\). In the case of 
identical distribution functions for electrons and positrons the {difference} in 
absorption of the O- and X-modes is proportional to $\tau \Delta N/N \approx \tau/\lambda$, 
that is almost negligible for $\lambda \gg 1$. It should be noted that the same result was 
obtained by Wang et al (2010) using exact solution of the equations for the Stokes 
parameters near the cyclotron resonance under the typical pulsar conditions. However, this 
result differs from one obtained by Petrova (2006).

Remember that for evaluation one can use the simple relation (Mikhailovsky et al. 1982)
\begin{equation}
\tau \approx \lambda (1 - \cos \theta_{\rm res}) \frac{r_{\rm res}}{R_{\rm L}}.
\label{taum}
\end{equation}
Hence, for $r_{\rm res} \approx 0.1 \, R_{\rm L}$, $\lambda \approx 10^{4}$, and 
$\theta_{\rm res} \approx 0.1$ the optical depth is to be high enough ($\tau \approx 10$).
On the other hand, as shown on Fig.~\ref{fig0} and Fig.~\ref{figres0}, for 
$r_{\rm res} \approx 0.1 \, R_{\rm L}$ the ray passes the very central parts
of the open field lines region where the plasma number density $n_{\rm e}$ can be 
much smaller than $\lambda n_{\rm GJ}$ ($g(f) \ll 1$). For this reason, as will
be shown below, the absorption of the outgoing radiation can be not so strong. 

{Finally, as it is shown on Fig.~\ref{figres}, in the case of rotating magnetosphere
${r}_{\rm{res}}$ depends on the pulsar rotation phase. Competition of these two effects (i.e., 
non-uniform plasma density distribution and difference in cyclotron radii) determines which 
part of the beam, leading or trailing one, will be absorbed more efficiently.}

\section{Limiting polarization}

The limiting polarization effect is well-known (Zheleznyakov 1977). When the radiation escapes 
into the region of rarefied plasma, the wave polarization ceases to depend on the orientation of 
the external magnetic field. At the same time, in the domain of the dense enough plasma where 
the geometrical optics approximation is valid, the orientation of the polarization ellipse is 
to be determined by the direction of the external magnetic field. This implies that the 
geometrical optics approximation under weak anisotropy conditions becomes inapplicable and the 
question about the pattern of the limiting polarization effect should be solved by using the 
equations that describe a linear interaction of waves in an inhomogeneous magnetoactive plasma.

Traditionally to describe general radiative transfer in magnetoactive plasma four first-order 
differential equations (for all four Stokes parameters) are used (Sazonov 1969; Zheleznyakov 
1996; Petrova \& Lyubarskii 1990; Broderick \& Blandford 2010; Wang et al. 2010; Shcherbakov 
\& Huang 2011). Budden eqution, i.e., the second-order equation to the complex function %${\cal V}$ 
actually corresponds to the same approach (Budden 1972; Zheleznyakov 1977). On the other hand, 
both the standard and the Zheleznyakov-Budden approaches are not quite {convenient} for quantitative 
estimates of the polarization of the escaping emission in general case. But since we are going to 
describe the propagation of originally fully polarized waves, not the {ensemble} of waves, we 
actually need only two equations for observable parameters, i.e., the position angle and the 
Stokes parameter $V$.  

There exists a different approach that allows us immediately write down the equations for these 
observable quantities, namely, the Stokes parameter $V$, defining the circular polarization  and 
the position angle $p.a.$, characterizing the orientation of polarization ellipse (Kravtsov \& 
Orlov 1990). This approach is valid in the quasi-isotropic case, i.e., in the case when the 
dielectric tensor can be presented as
\begin{equation}
\varepsilon_{ij} = \varepsilon \delta_{ij} + \chi_{ij},
\end{equation}
where the anisotropic part $\chi_{ij}$ is small as compared to isotropic one. In this case we 
have two small parameters --- general WKB parameter $1/kL$ and
\begin{equation}
\Delta n/n_{1,2}\sim\chi_{ij}/n_{1,2} \ll 1.
\label{vozm}
\end{equation}
As a result, the solution can be found by expansion over this two small parameters. 

As one can check, these conditions are just realized in the pulsar magnetosphere 
(Andrianov \& Beskin 2010). Indeed, in the region $r \sim r_{\rm esc} \sim 10^{3}R$ 
the value of $v = \omega_{\rm p}^2/\omega^2$ 
\begin{equation}
v \sim 10^{-7} \,  \lambda_{4} \, B_{12} \, \nu_{\rm GHz}^{-2} \, P^{-1},
%\left(\frac{r_{\rm esc}}{10^{3} R}\right)^{-3},
\end{equation}
is much smaller than unity. Accordingly, the deviation of the refractive indices from unity, 
$|n_{1,2}-1| \sim v$, is also very small here, so we can neglect the wave refraction in the 
polarization formation region.

The Kravtsov-Orlov equation 
\begin{eqnarray}
\frac{{\rm d}\Theta}{{\rm d}l} =\kappa+\frac{i\omega}{4c}
[(\chi_{ba}-\chi_{ab})
+(\chi_{ba}+\chi_{ab})\cos2\Theta-\nonumber \\(\chi_{aa}-\chi_{bb})\sin2\Theta],
\end{eqnarray}
is the equation for the complex angle $\Theta = \Theta_1 + i \Theta_2$,  where $\Theta_1$ 
is a position angle and $\Theta_2$ determines the circular polarization by the relation 
\begin{equation}
V = I \, {\rm tanh}2\Theta_2. 
\end{equation}
Here $I$ is the intensity of the wave. The components of the dielectric tensor $\chi_{ij}$ are to 
be written in a frame of unitary vectors ${\bf a}$ and ${\bf b}$ in the picture plane 
where ${\bf a}$ is determined by the projection of the vector $\nabla \varepsilon$. Finally, 
\begin{equation}
\kappa = 1/2({\bf a}\cdot[\nabla , {\bf a}] + {\bf b}\cdot[\nabla, {\bf b}]) 
\end{equation}
is the ray torsion (see Kravtsov \& Orlov 1990 for more detail).
It can be easily understood that the rotation of position angle described by the ray torsion 
is fictious and describes only the rotation of coordinate system. As a result, we can write 
down 
\begin{eqnarray}
\frac{{\rm d}\Theta_1}{{\rm d}l} = &&
\frac{\omega}{2c}{\rm Im}\,[\varepsilon_{x'y'}]
\nonumber \\
&& -\frac{1}{2}\frac{\omega}{c} \Lambda\cos[2\Theta_1-2\beta_{B}(l)-2\delta(l)]{\rm sinh}2\Theta_2,
\label{t1}\\
\frac{{\rm d}\Theta_2}{{\rm d}l} = &&
\frac{1}{2}\frac{\omega}{c}\Lambda\sin[2\Theta_1 - 2\beta_{B}(l)-2\delta(l)] {\rm cosh}2\Theta_2.
\label{t2}
\end{eqnarray}
Here $l$ is a coordinate along the ray propagation, and the angle $\beta_{B}(l)$ defines the 
orientation of the external magnetic field in the picture plane ({defined as 
$\tan \beta_{B} = B_{Y}/B_{X}$, where $B_{X}$ and $B_{Y}$ are the components of the 
magnetic field vector in a XYZ system, see Fig.~\ref{fig2s}}). Further,
\begin{equation}
\Lambda=\mp\sqrt{({\rm Re}\,[\varepsilon_{x'y'}])^2+\left(\frac{\varepsilon_{x'x'}-\varepsilon_{y'y'}}{2}\right)^2},
\label{Lambda}
\end{equation}
where the signs correspond to the regions before/after the cyclotron resonance and
\begin{equation}
\tan(2\delta)= - \frac{2{\rm Re}\,[\varepsilon_{x'y'}]}{\varepsilon_{y'y'}-\varepsilon_{x'x'}}.
\end{equation}
Finally, $\varepsilon_{i'j'}$ are the components of plasma dielectric tensor in the frame  
where the $z$-axis directs along the wave propagation and the external magnetic field lies 
in the $xz$-plane (see Appendix C). {As was already stressed, the singularities at the 
cyclotron resonance in equations (\ref{t1})-(\ref{t2}) are absent due to averaging over 
wide particle energy distribution. For distribution function (\ref{distrfunc}) this averaging 
can be done analytically.} 

We would like to note that in these equations the circular polarization is defined as it is 
common in radio astronomy (positive $V$ corresponds to LHC polarization). Nonrelativistic version of 
the above equations is given in Czyz et al. (2007). {It should be mentioned that the 
equation for the Stokes vector evolution has been recently shown to be derived directly from 
the Kravtsov-Orlov quasi-isotropic approximation (Kravtsov \& Bieg 2008)}.

\begin{figure}
\includegraphics[scale=0.6]{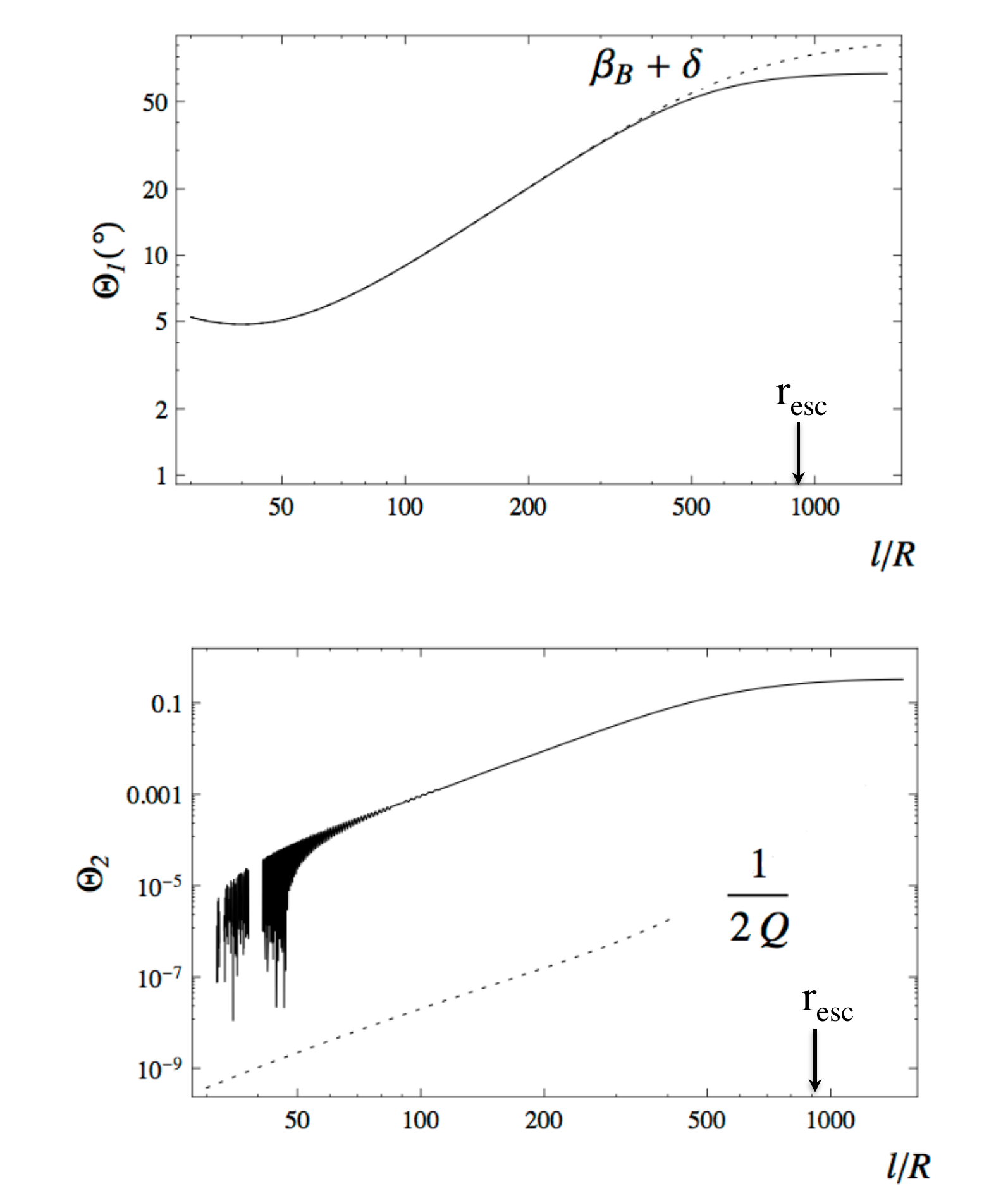}
\caption{Evolution of the angles $\Theta_{1,2}$ for O-mode, $\phi = -5^{\circ}$, 
$\lambda = \rm{10^3}$, $\gamma_{0} = \rm{50}$, $\nu = \rm{1 GHz}$, and magnetic field model B.
Their small {oscillations} in the region $r \ll r_{\rm esc}$ can be seen for $\Theta_2$ only
due to the different scales in the upper and lower panels. The dash lines correspond to
the geometrical optics values $\Theta_1 = \beta_{B} + \delta$ and $\Theta_2 = 1/2Q$}
\label{angles}
\end{figure}

As one can see on Fig.~\ref{angles}, in the geometrical optics region Eqns. (\ref{t1})--(\ref{t2}) 
describe {oscillations} of the angle $\Theta_{1}$ near the value $\Theta_{1} = \beta_{B} + \delta$. 
As the ray moves into the region of rarefied plasma, the length of the spatial {oscillations} 
$L \sim c/(\omega \Delta \, n)$ increases and in the region \mbox{$r > r_{\rm esc}$} becomes larger 
than the characteristic length $r$. As a result, the angles $\Theta_{1}$ and $\Theta_{2}$ 
become constant for $r \gg r_{\rm esc}$. They are the values that  characterize the outgoing 
radiation.

Thus, the basic equations (\ref{t1})-(\ref{t2}) generalize ones obtained by Andrianov 
\& Beskin (2010) for zero drift velocity ${\bf U}=0$ when ${\rm Re}\,[\varepsilon_{x'y'}] = 0$
and, hence, $\delta = 0$. In particular, they now include into consideration the aberration effect 
considered by Blaskiewicz et al. (1991). This effect was also considered by Petrova \& Lyubarskii 
(2000), but for the infinite magnetic field only. It is important that in Eqns. (\ref{t1})-(\ref{t2}) 
the angle $\Theta_1$ is measured relative to the laboratory frame because these equations contain 
the difference between $\Theta_1$ and $\beta_{B}$ only. 

Equations above have the following important property. For homogeneous media ($\beta_{B} = $ const, 
$\varepsilon_{ij} =$ const) the parameters of polarization ellipse $\Theta_1$ and $\Theta_2$ remain 
constant if the following conditions are valid:
\begin{eqnarray}
\Theta_1 & = & \beta_{B}+\delta, \quad \quad \quad
{\rm sinh}2\Theta_2 = \frac{{\rm Im}\,[\varepsilon_{x'y'}]}{\Lambda} =  -\frac{1}{Q}, 
\label{modes1}\\
\Theta_1 & = & \beta_{B}+\delta + \pi/2, \quad  
{\rm sinh}2\Theta_2 = -\frac{{\rm Im}\,[\varepsilon_{x'y'}]}{\Lambda} = \frac{1}{Q}.
\label{modes2}
\end{eqnarray}
Here (see the definition of $\epsilon_{i'j'}$ in Appendix D)
\begin{equation}
Q = i \, \frac{\epsilon_{y'y'} - \epsilon_{x'x'}}{2\epsilon_{x'y'}}.
\end{equation}
This closely corresponds to the polarization of the two normal modes, the former corresponding 
to the O-mode, and the latter to the X-mode. In addition, the following 
important property holds: irrespective of the pattern of change in plasma density and magnetic 
field along the trajectory, if two modes were orthogonally polarized in the beginning 
($\Theta_1^{(1)} - \Theta_1^{(2)} = \pi/2$, $\Theta_2^{(1)} = - \Theta_2^{(2)}$), then this 
property will also be retain subsequently, including the region where the geometrical optics 
approximation breaks down.

{Finally, as was already mentioned, in the region $r \ll r_{\rm esc}$ one can put
${\rm d}\Theta_1/{\rm d} l \approx {\rm d}(\beta_{B} + \delta)/{\rm d}l$.} Hence, for high 
enough shear of the external magnetic field along the ray propagation when the derivative 
\mbox{${\rm d}(\beta_{B} + \delta)/{\rm d}x$} is high enough, the first term in the r.h.s. of Eqn. 
(\ref{t1}) may be neglected. As for $\Theta_2 \ll 1$ we have 
${\rm sinh}2\Theta_2 \approx {\rm tanh}2\Theta_2$, one can write down
for $V/I = {\tanh}2\Theta_2$
\begin{eqnarray}
\frac{V}{I} \approx \frac{1}{|Q|} \frac{{\rm d}(\beta_{B} + \delta)/{\rm d}x}{A}
\frac{1}{\cos[2\Theta_1 - 2\beta_B(l) - 2\delta(l)]}.
\label{twom}
\end{eqnarray}
Here
\begin{eqnarray}
A = \left|
v_{\parallel}/c(1 - \sin \theta \, U_x/c)-\cos \theta (1 - U^2/c^2)\right|,
\label{V}
\end{eqnarray}
$x=\Omega l/c$, and we used $\lambda n_{GJ}$ for the plasma number density. 

Thus, the sign of the circular polarization will coincide with the sign of the 
derivative ${\rm d}(\beta_{B}+\delta)/{\rm d}x$ for the O-mode and they must be opposite 
for the X-mode. This approximation can be used for large enough derivative
\mbox{${\rm d}(\beta_{B}+\delta)/{\rm d}x \sim 1$} (i.e., for large enough total turn 
$\Delta (\beta_{B}+\delta) \sim 1$ within the light cylinder $R_{\rm L} = c/\Omega$), 
and for small angle of propagation $\theta \ll 1$ through the relativistic plasma 
($v_{\parallel}/c \sim 1$). Both these conditions are valid in the magnetospheres 
of radio pulsars  with a good accuracy. Indeed, assuming that $U/c \ll 1$ and 
$U_x/c \approx U/c \approx \theta$ one can obtain
\begin{equation}
A \approx \frac{\theta^2}{2} - \frac{1}{2\gamma^2} - \frac{U_x}{c}\sin \theta + \frac{U^2}{c^2} 
\approx \frac{\theta^2}{2} - \frac{1}{2\gamma^2} \ll 1.
\end{equation}
So, the Stokes parameter $V$ (\ref{twom}) is to be much larger than $V_0 = \pm I/Q$ resulting 
from standard evaluation (Ginzburg, 1961). 

{It is important that in this case in the region where the geometrical optics is valid the circular 
polarization is to be determined by the value of $\Lambda$ (\ref{Lambda}) which does not depend on 
imaginary non-diagonal components of the dielectric tensor  $\chi_{ab}$ and $\chi_{ba}$}. This 
fundamental property is well-known in plasma physics and crystal optics (see, e.g., Zheleznyakov 
et al. 1983; Czyz et al. 2007), but up to now it was not used in connection with the pulsar 
radio emission. {For radio pulsars this property is especially important because for 
electron-positron plasma the imaginary part of the dielectric tensor depends significantly on
the difference in particle energy distributions which is not known with the enough accuracy.}

Finally, our numerical simulations show that the sign of the derivative 
${\rm d}(\beta_{B}+\delta)/{\rm d}x$  is opposite to the sign of ${\rm d}p.a./{\rm d}\phi$. 
As one can see from Eqn. (\ref{twom}), this results in an important prediction:
\begin{itemize}
\item 
For the X-mode the signs of the circular polarization $V$ and the derivative 
${\rm d} p.a./{\rm d}\phi$ should be the SAME.
\item 
For the O-mode the signs of the circular polarization $V$ and the derivative 
${\rm d} p.a./{\rm d} \phi$ should be OPPOSITE.
\end{itemize}
This implies also that the effects of the particle drift motion, as was already 
found by Blaskiewicz et al. (1991) (see also Hibschman \& Arons 2001), shifts 
the {\it p.a.} curve to the trailing part of the mean profile. {As it is 
shown below, our results are in qualitative agreement with this statement. 
Moreover, at present there are some observational confirmations of this property 
(Mitra \& Rankin, 2011). Nevertheless, the main distinction of our theory is in 
self-consistent definition of $r_{\rm{esc}}$  (and, i.e., $p.a.$ shift value) on 
the direct solution of polarization transfer equations, that depends not only on 
the geometry, but on plasma parameters as well.}

{Typical evolution of angles $\Theta_1$ and $\Theta_2$ are presented on Fig.~\ref{angles}.
It shows that the analytical estimate of the escape radius (\ref{resc}) is qualitatively correct. 
Nevertheless, it should be mentioned that it depends on the plasma multiplicity factor, that 
depends effectively on pulsar rotation phase due to non-uniform plasma number density 
distribution. It also shows that real values of $\Theta_2$ are indeed much larger than 
the corresponding standard value $1/2Q$.}

\section{Results}

Thus, in this paper the arbitrary non-dipole magnetic field configuration, arbitrary number density profile
within the polar cap, the drift motion of plasma particles, and their realistic energy 
distribution function are taken into account. It gives us the first opportunity to provide 
the quantitative comparison of the theoretical predictions with observational data. Using 
numerical integration we can now model the mean profiles of radio pulsars and, hence, 
evaluate the physical parameters of the plasma flowing in the pulsar magnetosphere. 

It is necessary to stress that the detailed discussion of the morphological properties of 
mean profiles {resulting from different inclination and impact angles} is beyond the 
scope of our consideration {and are addressed to future papers}. The goal of this paper is 
in quantitative analysis of the propagation effects on the polarization characteristics of 
radio pulsars. In particular, we try to determine how the plasma parameters affect the 
$S$-shape of the position angle swing and the properties of the mean profile.

\subsection{Ordinary pulsars}

At first, let us discuss the results obtained by numerical integration of equations 
({\ref{t1})--({\ref{t2}) for ''ordinary'' pulsar (its parameters are given in 
Table~\ref{table1}). Everywhere below the dashed curves on the intensity panel 
show the intensity profile without any absorption. As was already stressed, in
this paper for simplicity we suppose that it repeats the particle number density profile 
shown in Fig.~\ref{fig0}. If the dashed curve is not shown, then the absorption is fatal 
and only the original intensity (which is normalized to 100 in its maximum) is shown. The 
dashed curves on {\it p.a.} panels show the prediction of the RVM-model (\ref{p.a.}). 
Finally, pulsar phase $\phi$ is measured in degrees everywhere below.

On Fig.~\ref{figres1} we show the intensity $I_{\infty}$ (\ref{intens}) (left panel) and
the {\it p.a.} swing (right panel) for extraordinary X-mode as a function of the pulsar 
phase $\phi$ for ''non-rotating dipole'' without the wind component (model A); the drift 
effects are neglected as well. The circular polarization degree does not exceed one percent 
here and that is why this curve is not presented in this picture. It results from 
approximately constant $\beta_{B}$ along the ray. For this reason, as we see, the {\it p.a.} 
curve is nicely fitting by the RVM model. 

Further, the upper solid line corresponds to $f_0 = 0.25$, and the lower one corresponds to 
$f_0 = 0.0025$. The lower intensity curve shows that for the very small core in the number 
density $f_0 = 0.0025$ ($r_{\perp}/R_0 = 0.05$) the absorption is fatal and the emission 
cannot escape from the magnetosphere. As was already stressed, this property can be 
easily explained. Indeed, for $f_0 \ll 1$ the rarefied region of the ''hollow cone'' is 
actually absent, and the rays pass the cyclotron resonance in the region of rather dense 
plasma. On the other hand, for $f_0 \approx 1$ the number density in the region of the 
cyclotron resonance is low enough for rays to escape the magnetosphere without strong 
absorption.

\begin{table}
\caption{Parameters of the 'ordinary' pulsar}  
\centering
\begin{tabular}{|c|c|c|c|c|c|c|c|}
  \hline
$P$   & $B_{0}$      & $\alpha$  & $\beta$ & $f_{0}$  & $r_{\rm em}$ & $\gamma_0$ & $\lambda$ \\
  \hline
$1$ s & $10^{12}$ G & $45^{\circ}$ & $-3^{\circ}$ & 0.25 &$30 R$  & 50 & $10^3$ \\
  \hline
\end{tabular}
\label{table1} 
\end{table}

\begin{figure*}
\includegraphics[scale=0.6]{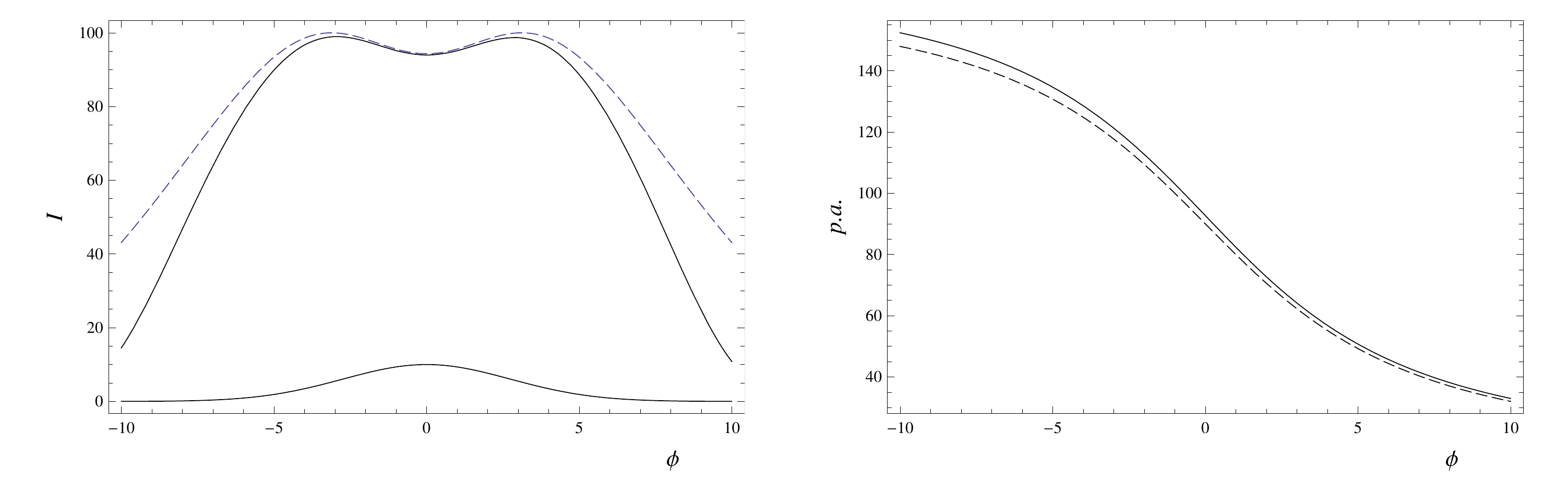}
\caption{The intensity $I_{\infty}$ (\ref{intens}) (left panel) and the {\it p.a.} swing 
(right panel) as functions of the pulsar phase $\phi$ (in degrees) for "non-rotating dipole" (model A). 
Here and below the dashed curves on the intensity panel show the intensity profile without 
any absorption. The upper solid line corresponds to \mbox{$f_0 = 0.25$,} and the lower one 
-- to $f_0 = 0.0025$} 
\label{figres1}
\end{figure*}

\begin{figure*}
\includegraphics[scale=0.28]{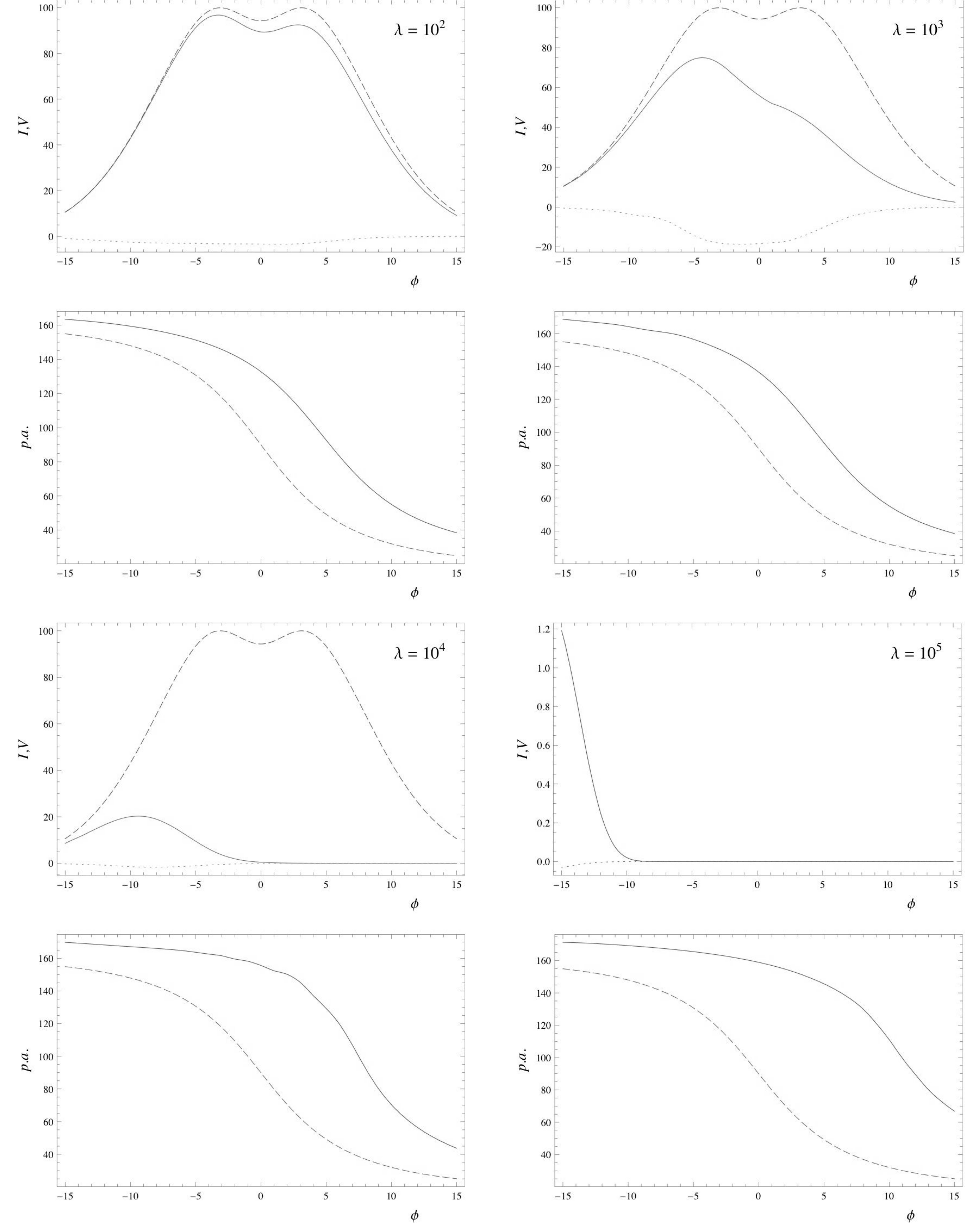}
\caption{The same for model C and for different multiplicity factors $\lambda = 10^2$, 
$10^3$, $10^4$, and $10^5$. Here $\gamma_0$ = 50 and $\nu = \rm{1GHz}$. Dotted lines correspond to
Stokes parameter $V$}
\label{figres2}
\end{figure*}

\begin{figure*}
\includegraphics[scale=0.28]{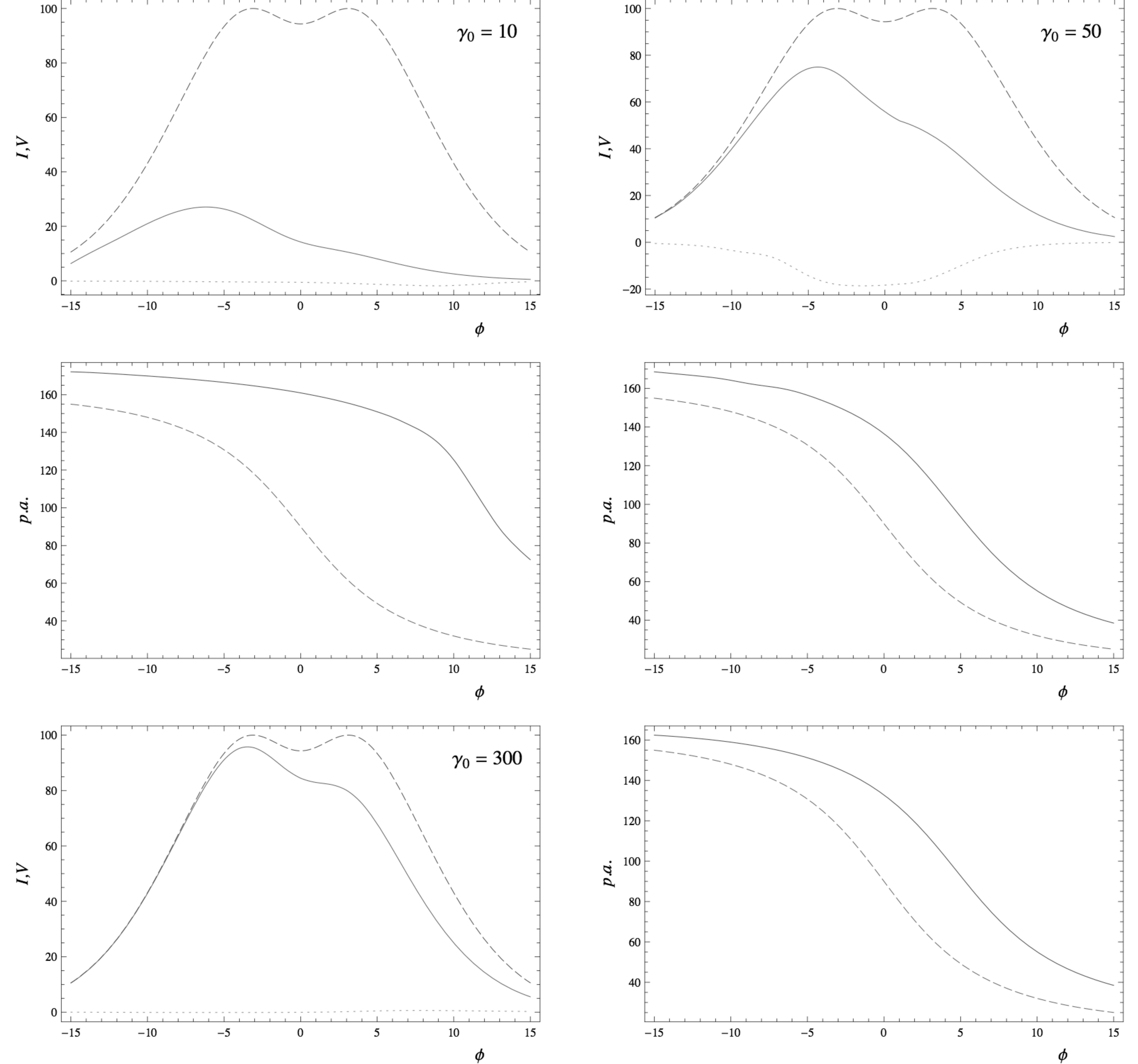}
\caption{The same for various Lorentz-factors $\gamma_0 = 10$, $50$, and $300$. Here  
$\lambda = 10^3$, and $\nu = 1 \rm{GHz}$} 
\label{figres3}
\end{figure*}

\begin{figure*}
\includegraphics[scale=0.28]{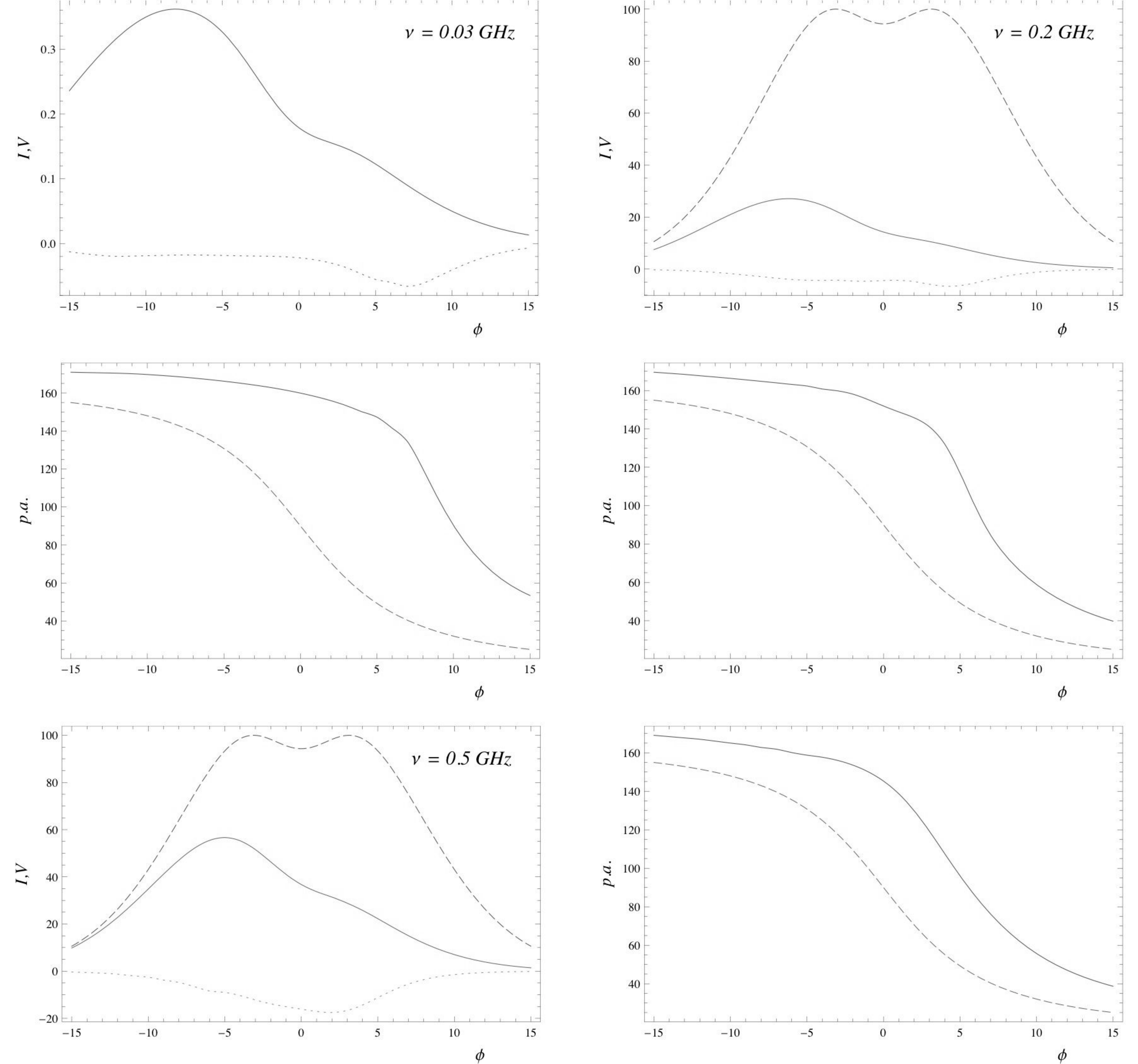}
\caption{The same for model C and for various frequencies $\nu = 0.03$, $0.2$, and $0.5$ GHz. 
Here $\lambda = 10^3$, and $\gamma_0 = \rm{50}$ } 
\label{figres4}
\end{figure*}

\begin{figure*}
\includegraphics[scale=0.6]{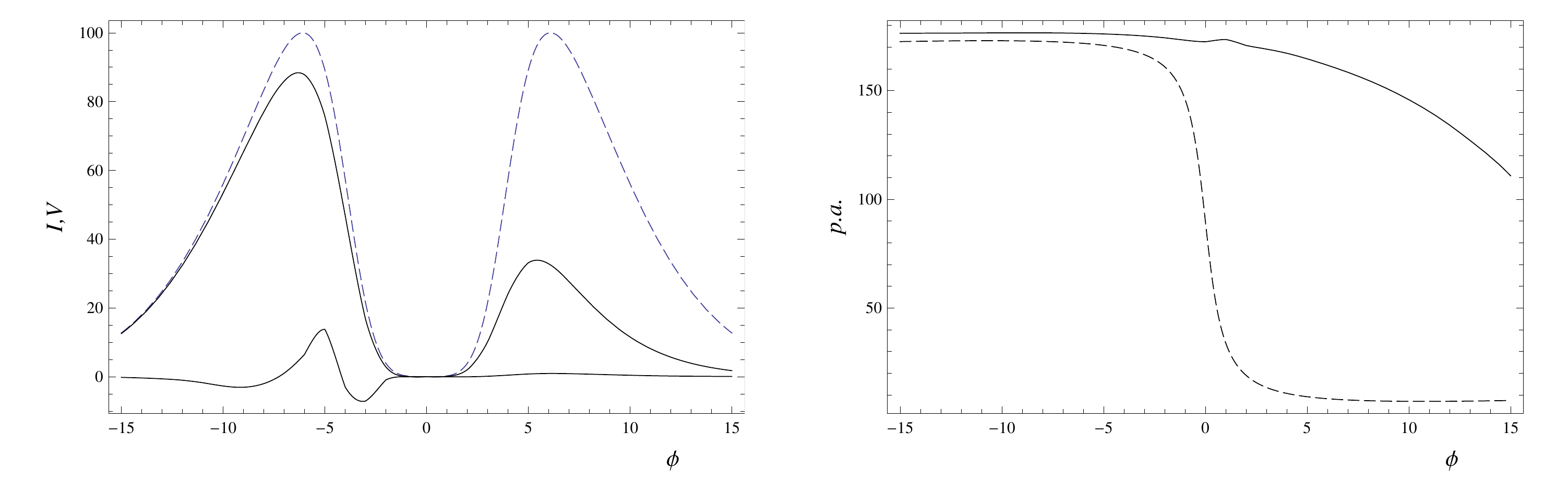}
\caption{The results of simulations for X-mode, $P =1s$, $\beta = -0.5^{\circ}$, 
$\alpha = 48.5^{\circ}$, $\lambda = 10^3$, $\gamma_0 = 50$, and $r_{\rm em}$ = 100$R$. It shows 
that under certain conditions the change of $V$ sign in the same mode may occur} 
\label{core}
\end{figure*}

{Below we consider model C as the main magnetic field model of our simulation. 
As to model B, one can find the examples of mean profiles basing on this model in Wang 
et al. (2010). The main differences between our results and ones presented in the  
paper mentioned above are in shifting of the $p.a.$ curve to the trailing part 
(because of the particle drift motion) and in possibility of absorption of the leading 
part of the pulse (due to the non-zero toroidal magnetic field).} 

On Fig.~\ref{figres2} we show the {\it p.a.} swing (lower panels), the intensity $I_{\infty}$ 
(solid lines on the top panels), and the Stokes parameter $V$ (dotted lines) as a function
of the pulsar phase $\phi$ for extraordinary wave for magnetic field model C and for various 
multiplicity parameter $\lambda$. It is obvious that the absorption increases with increasing 
$\lambda$. In most cases, the trailing part of the mean profile is absorbed (see Dyks et al.
2010 as well).

Besides, as one can see, the {\it p.a.} curves differ significantly from the RVM one 
(dashed lines) as \(\lambda\) growing. This property can be easily understood as well. Indeed, 
as the escape radius $r_{\rm esc}$ (\ref{resc}) increases as $\lambda^{2/5}$, for large enough 
$\lambda$ the polarization properties of the outgoing waves are to be formed in the vicinity of 
the light cylinder, i.e., in the region with quasi-homogeneous magnetic field 
(see Fig.~\ref{figmagnfield}).

Further, as was already mentioned, the drift effect causes the {\it p.a.} 
curve to be shifted to the trailing part of the mean profile. It is necessary to stress that the 
opposite shift is to take place if we neglect the drift effect on the dielectric tensor (Andrianov 
\& Beskin 2010; Wang et al. 2010). One can note that for high values of multiplicity parameter 
(i.e., for full absorption of the trailing part of the mean pulse) the observer will detect 
approximately constant {\it p.a.} Finally, as one can see from Eqn. (\ref{V}), the maximum 
of circular polarization is also shifted to the trailing side, as larger deviations from the 
$S$-shape produce larger circular polarization. It is not visible on this picture because the 
trailing side of the beam is absorbed. {Thus, one can conclude that the self-consistent 
quantitative analysis of observational data is to include these effects into consideration.}

{Another point should be mentioned. It is known that correct accounting of the inverse 
Compton scattering of the surface X-rays can reduce the pair multiplicity significantly (see, 
e.g., Hibschman \& Arons 2001). In this case plasmas effects on emission polarization can be 
negligible.}

On Fig.~\ref{figres3} we show the same dependences for various Lorentz-factors of outgoing plasma
$\gamma_0 = 10$, $50$, and $300$. As the escape radius $r_{\rm esc}$ (\ref{resc}) decreases 
as $\gamma^{-6/5}$, the largest shift of the $p.a.$ curve takes place for small $\gamma_0 = 10$.
Finally, on Fig. \ref{figres4} one can see the same dependences for various wave frequencies 
$\nu= 0.03$, $0.2$, and $0.5$ GHz. As $r_{\rm esc} \propto \nu^{-2/5}$, the largest shift of the 
{\it p.a.} curve takes place for small frequencies. One can note that the full investigation of 
frequency dependence of mean profiles of radio pulsars is more complicate and must include, e.g., 
the detailed analysis of frequency dependence of the emission radius $r_{\rm em}$ (BGI). This is 
beyond the scope of the article.

As was demonstrated above, in general the sign of the circular polarization  
remains constant for a given mode. But under certain conditions the change of 
the $V$ sign in the same mode may occur. It can take place when we cross the 
directivity pattern in the very vicinity of the magnetic axis. Such an example 
for the X-mode is shown on Fig.~\ref{core} for $\mbox{P = 1s}$, $\zeta = 49^{\circ}$, 
$\alpha = 48.5^{\circ}$, $\lambda = 10^3$, $\gamma_{0} = 50$, and $r_{\rm em} = 100 \, R$. 
{Certainly, this example is illustrative only. To explain the change of the
sign of the circular polarization in the centre of the mean profile the separate
consideration is necessary.}

\begin{figure*}
\includegraphics[scale=0.6]{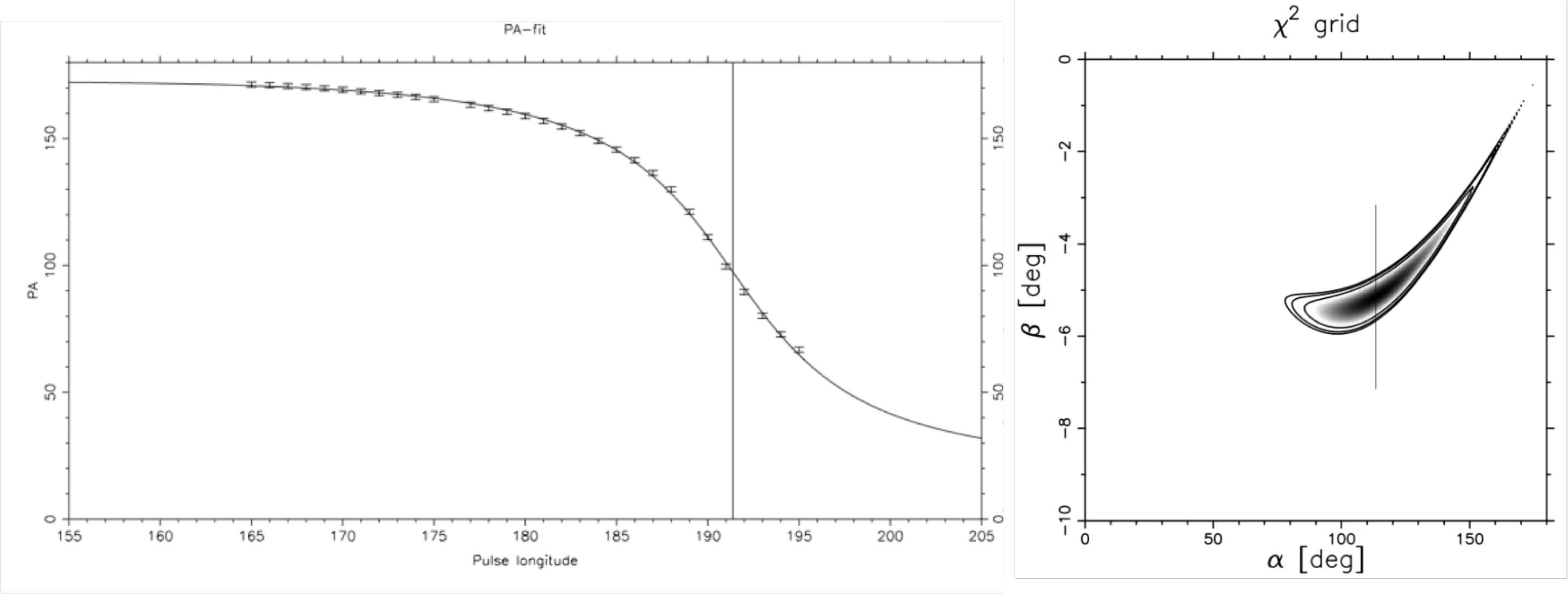}
\caption{The left panel shows the {\it p.a.} swing (points) for $\gamma_0 = 50$, $\lambda = 10^5$, 
$\nu = \rm{1GHz}$, $\alpha = 45^{\circ}$, and $\beta = -3^{\circ}$. Solid line corresponds to the best 
RVM fit giving unrealistic values  $\alpha = 97^{\circ}$ and $\beta = -5^{\circ}$. The right panel
shows $\chi^2$ map for RVM fitting}
\label{fit}
\end{figure*}

Finally, in Table~\ref{table2} we present the values of maximum derivative 
$({\rm d} p.a./{\rm d}\phi)_{\rm max}$, that is commonly used for 
determination of the inclination angle $\alpha$ (see, e.g., Kuzmin \& 
Dagkesamanskaya 1983; Malov 1990;  Everett \& Weisberg 2001). 
As we see, these values significantly depend on the plasma 
parameters (and can differ drastically from the RVM value). Hence,
more precise specifying of magnetospheric and plasma model is 
necessary for quantitative analysis of the pulsar characteristics. 

{In our opinion, this technique can be applied in future for radio 
pulsars for which the whole $S$-swing is detected. Otherwise, the situation 
like in Fig.~\ref{figres2} for $\lambda = 10^{4}$ may occur, when one can 
detect only constant part of the $ p.a.$ swing}. {At any way, one can 
conclude that if one try to fit the whole {\it p.a.} curve by the RVM function 
(\ref{p.a.}), then for low enough number density ($\lambda < 10^4$) and high 
enough particle energy ($\gamma_{0} > 50$) the angles $\alpha$ and $\beta$ 
obtained will satisfy the reality. But in other cases unexpected result may 
occur. As shown on Fig.~\ref{fit}, left panel (see Table~\ref{table1} as well), 
the simulated curve could be nicely fitted by the RVM model with non-zero 
shift value (\ref{p.a.}), but with the angles $\beta \approx -5^{\circ}$ and 
$\alpha \approx 97^{\circ}$ that drastically differ from real ones 
$\alpha = 45^{\circ}$ and $\beta = -3^{\circ}$. On the other hand,
the precision in determination of the angles $\alpha$ and $\beta$ 
may be not so high (see Fig.~\ref{fit}, right panel).}

\begin{table*}
\caption{The position angle maximum derivative $({\rm d}p.a./{\rm d}\phi)_{\rm max}$ and the 
shift $\Delta \phi$ of its position}  
\centering
 \begin{tabular}{|c|c|c|c|c|c|}
  \hline
$\lambda$& $\gamma_{0}$& $\nu$ (GHz)&$({\rm d}p.a./{\rm d}\phi)_{\rm max}$& ${\rm RVM}$ & $\Delta \phi (^{\circ})$  \\
\hline
$10^2$ &  50 & 1    &  $-9.47$ & $-10.14$ &  4.7 \\
\hline\
$10^4$ &  50 & 1    & $-14.47$ & $-10.14$ &  7.4 \\
\hline
$10^5$ &  50 & 1    & $-11.70$ & $-10.14$ & 10.5 \\
\hline
$10^3$ &  10 & 1    & $-12.72$ & $-10.14$ & 11.5 \\
\hline
$10^3$ &  50 & 1    &  $-9.86$ & $-10.14$ & 4.3 \\
\hline
$10^3$ & 100 & 1    &  $-9.47$ & $-10.14$ & 4.7 \\
\hline
$10^3$ & 300 & 1    &  $-9.46$ & $-10.14$ & 4.0 \\
\hline
$10^3$ &  50 & 0.03 & $-15.92$ & $-10.14$ & 8.4 \\
\hline
$10^3$ &  50 & 0.5  & $-11.82$ & $-10.14$ & 3.8 \\
\hline
$10^3$ &  50 & 0.2  & $-18.02$ & $-10.14$ & 5.5 \\
\hline
\end{tabular}
\label{table2} 
\end{table*}

\subsection{Two modes profiles}

\begin{figure*}
\includegraphics[scale=0.65]{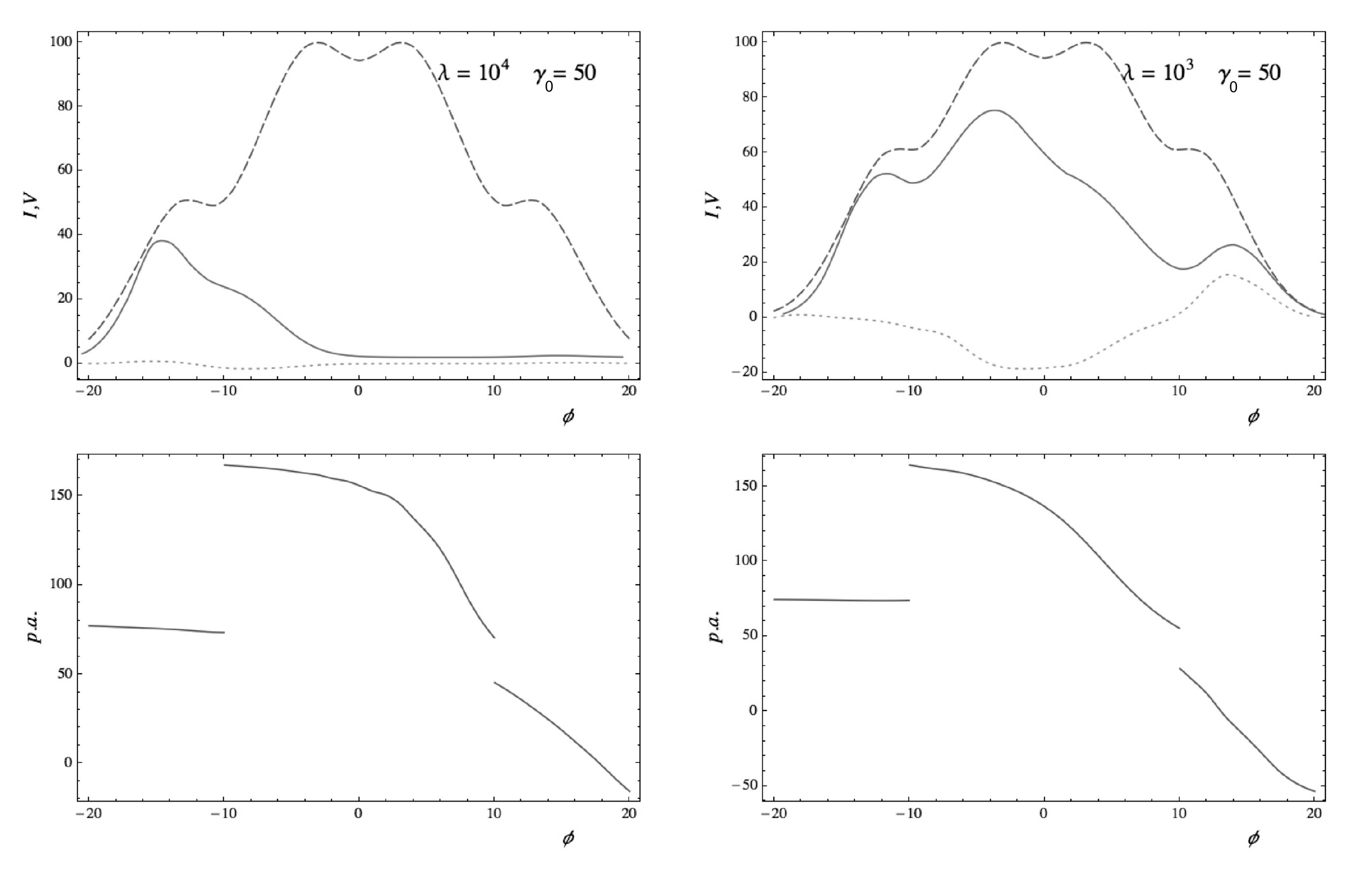}
\caption{Main profiles including both modes (O and X ones) for $\lambda = 10^{3}$ and $10^{4}$. 
Here $\gamma_{0} = 50$, and $\nu = \rm{1GHz}$} 
\label{2modes}
\end{figure*}

On Fig.~\ref{2modes} an examples of the mean profiles including two orthogonal modes 
are presented. The ratio of the intensities in the radiation domain $r = r_{\rm em}$ 
was assumed to be \mbox{$I_{\rm O}^{(0)}/I_{\rm X}^{(0)}=1/3$.} The jumps in the 
{\it p.a.} curves were done at the phase $\phi$ where $I_{\rm O} = I_{\rm X}$. As we 
see, this jump can differ from $90^{\circ}$. This results from the different trajectories 
of two orthogonal modes. It is necessary to stress that in our consideration the ratio 
$I_{\rm O}^{(0)}/I_{\rm X}^{(0)}$ is free, and its precise determination is a question 
for the radio emission generation theory. 

We see that for radio pulsars for which two modes can be observed the mean profiles
are indeed to have the triple form, which, in general, are not to be symmetric. As 
the O-mode deviates from magnetic axis, we have to see the O-mode in the leading part, 
the X-mode in the centre, and again the O-mode in the trailing part of the mean profile. 
{The detailed comparison with observational data will be 
prepared in the separate paper.}

%The detailed analysis is beyond the scope of our present consideration.   
 
\subsection{Millisecond pulsars}

As the last example, on Fig.~\ref{milly} we show the mean profiles obtained for millisecond 
pulsar (the parameters are given in Table~\ref{table3}). One important feature appearing 
here is that the leading, not trailing part of the mean profile can be absorbed. It is 
caused by the bending of the open field lines tube near the light cylinder due to non-dipolar 
magnetic field. As a result, the cyclotron resonance takes place in the region of rarefired 
plasma for the trailing part of the pulse. Also, stronger deviations from the $S$-shape of 
{\it p.a.} swing are found as compared to the case of ordinary pulsar, because the polarization 
forms closer to the light cylinder.

\begin{figure*}
\includegraphics[scale=0.28]{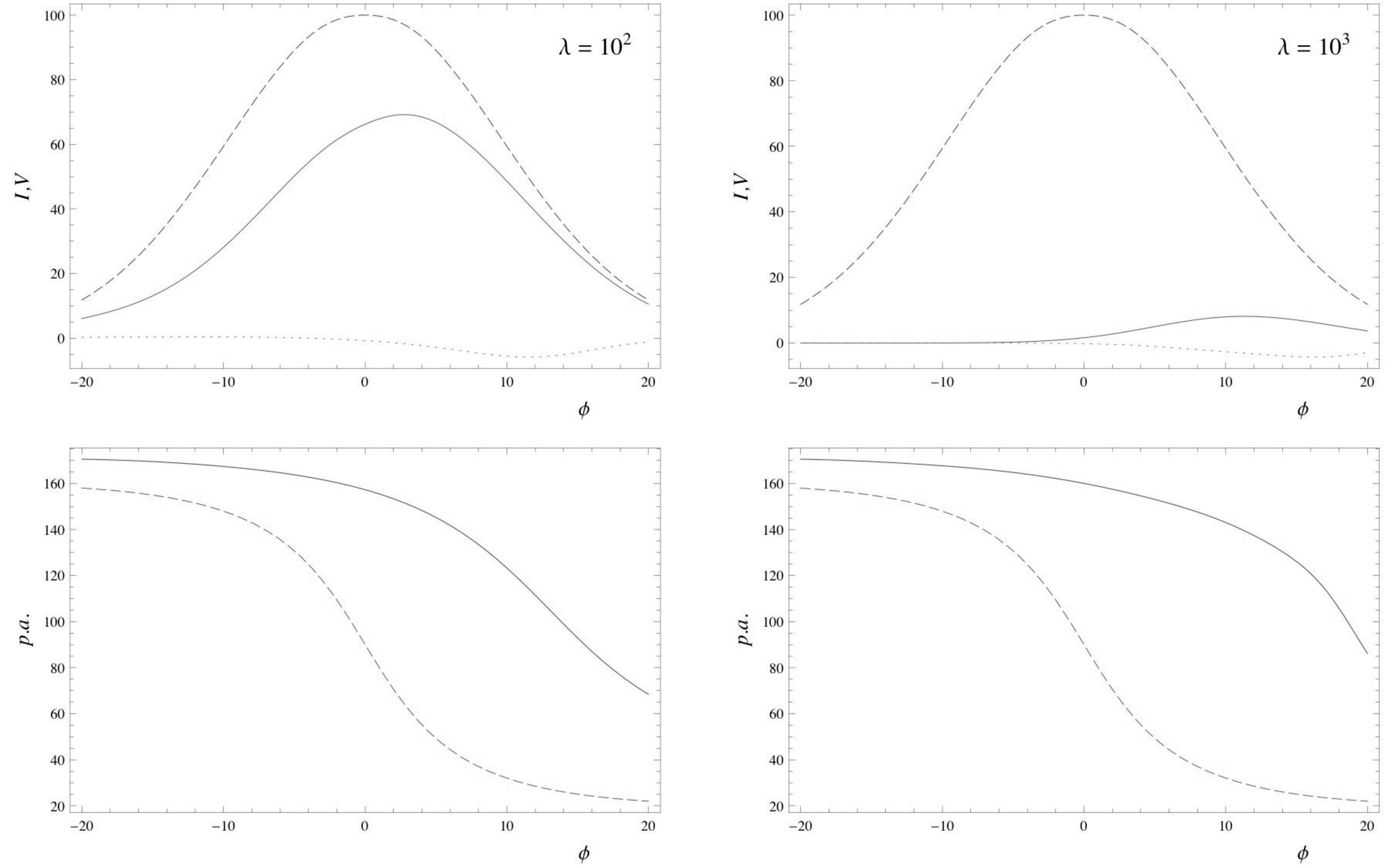}
\caption{The results of simulations for X-mode for the millisecond pulsar for multiplicity 
parameters $\lambda = 10^2$ (left panel) and $\lambda = 10^3$ (rigth panel)} 
\label{milly}
\end{figure*}

\begin{table}
\caption{Parameters of the millisecond pulsar}  
\centering
\begin{tabular}{|c|c|c|c|c|c|c|c|}
  \hline
$P$   & $B_{0}$      & $\alpha$  & $\beta$ & $f_{0}$  & $r_{\rm em}$ & $\gamma_{0}$ & $\lambda$ \\
  \hline
$20$ ms & $10^{8}$ G & $45^{\circ}$ & $-3^{\circ}$ & 0.04 &$1.25 R$  & 50 & $10^3$ \\
  \hline
\end{tabular}
\label{table3} 
\end{table}

\section{Discussion and conclusions}

In this paper we study the influence of the propagation effects on the mean profiles of 
radio pulsars. The Kravtsov-Orlov approach allows us firstly to include into consideration 
the transition from geometrical optics to vacuum propagation, the cyclotron absorption, and 
the wave refraction simultaneously. Arbitrary non-dipole magnetic field configuration, drift 
motion of plasma particles, and their realistic energy distribution were taken into account. 
Using numerical integration, {we found how the propagation effects can correct the
main characteristics of the hollow cone model.}

\begin{figure*}
\includegraphics[scale=0.32]{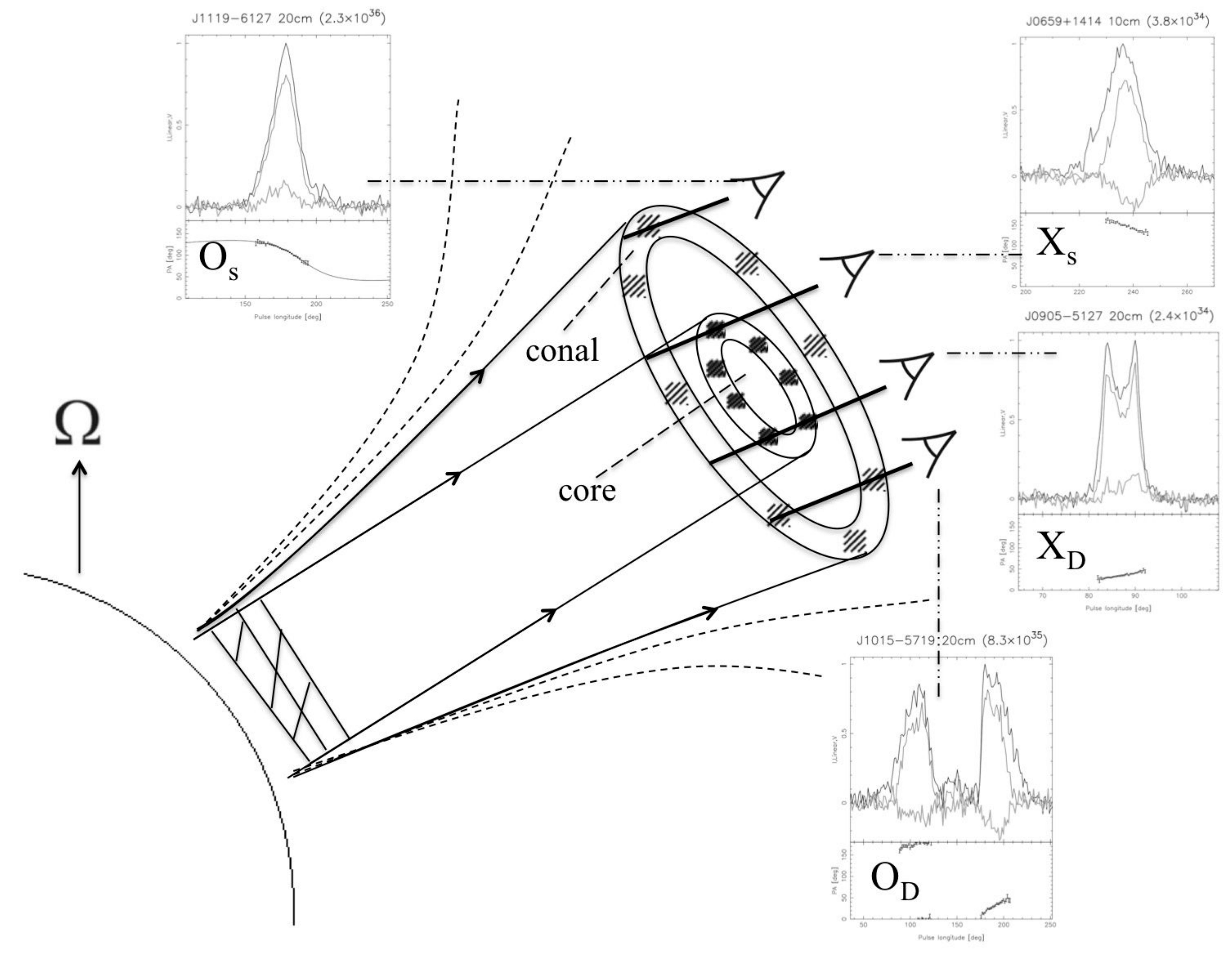}
\caption{Formation of the directivity pattern in the pulsar magnetosphere. The X-mode propagating rectilinearly
forms the core component, and deflected from the magnetic axis O-mode forms the conal one. The lower curves on each 
top panels indicate the Stokes parameter $V$, and the bottom panels shows the $p.a.$ swing. Corresponding pulsar
profiles was taken from Weltevrede \& Johnston (2008)}
\label{CCmod}
\end{figure*}

To summarize, one can formulate the following results:
\begin{itemize}
\item
We confirm the one-to-one correlation between the signs of circular polarization $V$ and 
the position angle derivative (${\rm d} p.a./{\rm d} \phi$) along the profile for both ordinary 
and extraordinary waves. For the X-mode the signs should be the SAME, and OPPOSITE 
for the O-mode. 
\item
The standard $S$-shape form of the ${\it p.a.}$ swing (1) can be realized for small enough 
multiplicity $\lambda$ and large enough bulk Lorentz-factor $\gamma$ only. In other cases 
the significant differences can take place. The location of the $p.a.$ maximum derivative 
$({\rm d}p.a./{\rm d}\phi)_{\rm max}$ is shifted to the trailing side of the pulse. 
\item
The value of $p.a.$ maximum derivative $({\rm d}p.a./{\rm d}\phi)_{\rm max}$, that is often 
used for determination the angle between magnetic dipole and rotation axis, depends on the 
plasma parameters [and differs from rotation vector model (RVM) value] and, hence, cannot 
be used without more precise specifying of magnetospheric plasma model.
\item
In general, the trailing side of the emission beam is absorbed. 
\end{itemize}

{In our opinion, even the preliminary consideration demonstrates that these conclusions are 
not in contradiction with observational data. The first statement is in good agreement with observations 
(Han et al. 1998;  Andrianov \& Beskin 2010). In some cases, the sign reversal in the core can occur 
(see Fig.~\ref{core} and Wang et al. 2010), that is detected for several pulsars (see also Radhakrishnan 
\& Rankin 1990). The second point is in agreement with empirical model of Blaskiewicz et al. (1991). On 
the other hand, as the shift value depends on plasma parameters (assuming fixed geometry), the solution 
of the inverse problem (i.e., the fitting of the real pulsar profiles) can in principle provide us a chance  
of correct enough estimating of plasma parameters. Finally, the last point is in agreement with observations 
as well (see, e.g., Backus et el., 2010). In more detail, it takes place for large enough multiplicity parameter 
$\lambda > 10^{3}$. But in some cases the leading part can be absorbed as well. It happens when the polarization 
forms close to the light cylinder.}

{Finally, Fig.~\ref{CCmod} demonstrates our understanding of the X- and O-modes
propagation in the pulsar magnetosphere. Here the dot-dash lines going to the panels 
with pulsar profiles show different intersections of directivity pattern, that form the 
observable profiles. Resulting from the different propagation, the O- and X-modes produce 
two concentric cones, the inner (core) part corresponding to the X-mode and the outer 
(conal) one to the O-mode. Remember that the analysis of the relative intensity in O- and 
X-modes is a question for generation theory. In the context of the present propagation 
theory this parameter should be considered as a free one.}

{Another source of uncertainty which can affect the formation of the mean profile
(and, in particular, results in the formation of its more complicated shape) is the
possible ''spotty'' of the directivity pattern, i.e., the presence of the separate
radiative domains within the diagram (Rankin 1993). As is well-known, such separate
domains are observed as a subpulse drift (Ruderman \& Sutherland 1975, Deshpande \&
Rankin 2001).}

{In addition, the full analysis is to include the possible linear
depolarization. E.g., depolarization at the edges of the profile can be easily
interpreted as a mixing of X and O-modes (Rankin \& Ramachandran 2003). But the
total degree of linear polarization (e.g., if we clearly see only one mode) is
a question to propagation in the generation region. If this region is rather
thick then one can expect the possible depolarization. For this reason we do
not treat here the total degree of linear polarization at all.}

As we see, in general if only one mode dominates in pulsar profile, then we expect 
statistically $D$ (double) mean profile for the O-mode and $S$ (single) one for the X-mode. 
Of course, we suppose here that there is nonzero natural width of the radiative domains that 
efficiently smooths the hollow cone corresponding to the narrow X-mode. Nevertheless, as is 
shown on Fig.~\ref{CCmod}, the X-mode can in principal produce $D$ profile and O-mode can 
produce $S$ ones. {The $X_{D}$ and $X_{S}$ profiles can be obtained when only X-mode 
is produced in the formation region}. On the other hand, if there are two modes with the similar 
intensity, the triple (T) and multiple (M) profiles can be obtained as well. So, this picture 
is in agreement with Rankin's morphological classification (Rankin 1983). Our investigation 
provides an additional information to morphological structure, i.e., the information of the type 
of the mode that forms mainly the mean profile. As was demonstrated, only the correlation 
between $p.a.$ derivative and $V$ signs provides us this chance.

{Finally, the determination of the angular dimension of the core and conal beams depends 
greatly on the emission radius (i.e., again, on the generation mechanism). So the detailed 
analysis is also beyond the scope of current investigation (one can find empirical results 
on this in, e.g., Maciesiak \& Gil 2011). 

Thus, the approach developed in this paper allows the predictions of the theory of 
radio emission to be quantitatively compared with the observational data. 
Moreover, based on the observational data, one can reject a number of models in which 
the shapes of the profiles of the position angle and the degree of circular polarization 
never realized in practice are obtained. Thus, it becomes possible to solve the inverse 
problem, i.e., to determine the parameters of the outflowing plasma and the magnetic 
field structure.

We will be glad to any collaboration with observers for investigating the 
observational features of the mean profiles on the base of our theory. The source 
code of our calculations (in Mathematica 7 or C) is available under request.

\section{Acknowledgments}

We thank Prof. A.V.~Gurevich and Ya.N.~Istomin for his interest and support, and J.~Dyks,
A. Jessner, P. Jaroenjittichai, M. Kramer, V.V. Kocharovsky, D.~Mitra, M.V.~Popov, R.~Shcherbakov, 
B.~Rudak, and H.-G.~Wang for useful discussions. We also thank A.~Spitkovsky for valuable 
opportunity and assistance in dealing with his numerical solution and for outstanding discussions. 
This work was partially supported by Russian Foundation for Basic Research (Grant no. 11-02-01021). 

\appendix

\section{Refraction}

\begin{figure}
\includegraphics[scale=0.3]{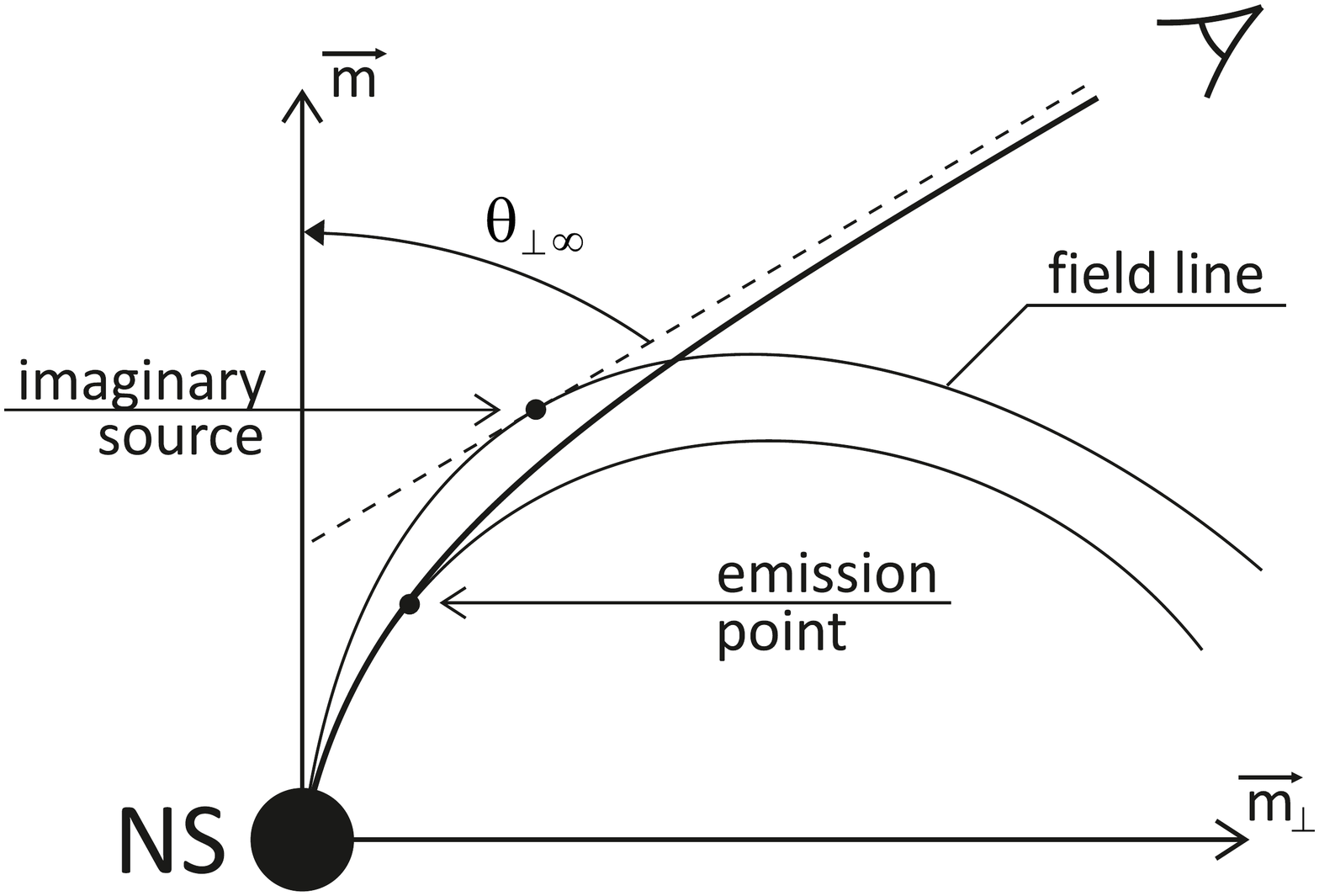}
\caption{Imaginary source corresponding to rectilinear propagation of the wave} 
\label{figRefract}
\end{figure}

In this work we consider the simple model of the refraction obtained under the 
assumption that the plasma number density is constant within the polar cap, i.e., 
$g(\theta_{m}, \varphi_{m}) = 1$. To include refraction into consideration
we introduce the ''imaginary source'' of  radiation giving the same trajectory 
at large distances whence emitted parallel to the magnetic field line (see 
Fig. \ref{figRefract}). 
%This fact describe $\cos \theta_{\perp\infty} = ({\bf m}(\phi),{\bf{e_z}})$ 
%for each pulsar phase. 
As the ''tearing off'' level locates deeply in the magnetosphere, i.e., 
$r_{\rm A} \ll R_{\rm L}$, one can use the analytical expression (\ref{angle})
for the angle $\theta_{\perp \infty}$ (BGI 1993). It gives for the polar angle 
of the emission point
\begin{equation}
\theta_{\rm{em}}\approx \left[\left(\frac{l_r}{R}\right)^{0.21}
\theta_{\perp\infty}\left(\frac{1}{\omega^2}
\left<\frac{\omega^2_{p0}}{\gamma^3}\right>\right)^{-0.07}\right]^{1.39}
\end{equation}
(for dipole magnetic field it does not depend on the radial distance).
As a result, we obtain for the polar angle of imaginary source
\begin{equation}
\tan \theta_{\rm{im}} = \frac{4}{3}\frac{\tan\theta_{\perp\infty}}{1+\sqrt{1+8/9\tan^2 
\theta_{\perp\infty}}}.
\end{equation}
After some algebraic calculations, one can find the trajectory
$$
{\bf R} = {\bf m}(\phi)\rho\cos \theta_{\rm{im}}  
+ {\bf m_{\perp}}(\phi)\rho\sin \theta_{\rm{im}} + r{\bf e_{z}}.
$$
Here
\begin{equation}
\rho(\phi) = 
 l_t\frac{|\sin \beta - 
    \cos \beta \tan \theta_{\perp\infty})|}{
  |\tan \theta_{\rm{im}} - \tan \theta_{\perp\infty}|}
  \sqrt{1 + \tan^2 \theta_{\rm{im}}}
\end{equation}
is the radius of imaginary source, and
\begin{eqnarray}
\beta = \frac{\Psi}{2} + \frac{1}{2}\, \arcsin\left(\frac{\sin \Psi}{3}\right),
\end{eqnarray}
where
\begin{equation}
\Psi = \theta_{\perp\infty} 
-\left[\frac{1}{\omega^2}<\frac{\omega^2_{p0}}{\gamma^3}>
\left(\frac{l_t}{R}\right)^{-3}\right]^{1/4}.
\end{equation}
Finally, \(\bf m_{\perp}(\phi)\) is a unit vector perpendicular to ${\bf m}(\phi)$ lying
for every pulsar phase \(\phi\) in the plane containing the magnetic ${\bf m}(\phi)$ and 
the wave ${\bf k}$ vectors.

\section{Derivation of dielectric tensor}

Since all the quantities in equation (\ref{euler}) are proportional to 
$\exp{(- i \omega t + i{\bf kr})}$, its solution can be easily obtained:

\begin{widetext}

\begin{eqnarray}
\delta v_x & = & \frac{1}{\omega^2_B - \gamma^2\gamma_{U}^2\tilde{\omega}^2}
\left[-i\tilde{\omega}\gamma \left(1 + \gamma_{U}^2\frac{U^2_{y}}{c^2}\right)D_1 
+ \left(\omega_B + i\tilde{\omega}\gamma\gamma_{U}^2\frac{U_xU_y}{c^2}\right)D_2\right],\\
\delta v_y & = & \frac{1}{\omega^2_B - \gamma^2\gamma_{U}^2\tilde{\omega}^2}
\left[-i\tilde{\omega}\gamma \left(1 + \gamma_{U}^2\frac{U^2_{x}}{c^2}\right)D_2 
+ \left( - \omega_B + i\tilde{\omega}\gamma\gamma_{U}^2\frac{U_xU_y}{c^2}\right)D_1\right],\\
\delta v_z & = & i\frac{e \gamma_{U}^2}{m_{\rm e}\omega\tilde{\omega}\gamma^3}
\left[\tilde{\omega}\delta E_z + k_z({\bf V_{0}}, \delta {\bf E} )\right]  
- \frac{v_{\parallel}}{c^2}\gamma_{U}^2(U_x\delta v_x + U_y\delta v_y),
\end{eqnarray}
where
\begin{eqnarray}
D_1 & = & \frac{e}{m_{\rm e}\omega}\left[\tilde{\omega}\delta E_x + k_x({\bf V_{0}}, \delta {\bf E} ) 
-  \gamma_{U}^2\frac{U_x v_{\parallel}}{c^2} 
\left(\tilde{\omega}\delta E_z +k_z({\bf V_{0}}, \delta {\bf E} )\right)\right],\\
D_2 & = & \frac{e}{m_{\rm e}\omega}\left[\tilde{\omega}\delta E_y  
- \gamma_{U}^2\frac{U_y v_{\parallel}}{c^2} 
\left[\tilde{\omega}\delta E_z + k_z({\bf V_{0}}, \delta {\bf E} )\right]\right].
\end{eqnarray}
\end{widetext}

From (\ref{continuity}) one can find the number density pertubation
\begin{equation}
\delta n_{\rm e} = \frac{n_{\rm e}}{\tilde{\omega}}\left({\bf k}, \delta {\bf v}\right).
\end{equation}
After making substitution to (\ref{cond}) we obtain for dielectric tensor components:

\begin{widetext}

\begin{eqnarray}
\varepsilon_{xx} & = & 1 - <\frac{\gamma_{U}^2 k_z^2 U^2_{x}\omega^2_p}{\tilde{\omega}^2\gamma^3\omega^2}> 
+ <\frac{\omega_{\rm p}^2[\tilde{\omega}_0^2+U_x^2/c^2 (\gamma_{U}^2 k_z^2 v_{\parallel}^2 -\omega^2)]
\gamma\gamma_{U}^2}{\omega^2(\omega_B^2-\gamma^2\gamma_{U}^2\tilde{\omega}^2)}>,\\
\varepsilon_{xy} & = & - <\frac{\gamma_{U}^2 k_z^2 U_{x}U_{y}\omega^2_p}
{\tilde{\omega}^2\gamma^3\omega^2}> 
+ i<\frac{\gamma_{U}^2 \omega_{\rm p}^2\omega_B(\tilde{\omega}_{0}-\omega U^2/c^2)}
{\omega^2[\omega_B^2-\gamma^2\gamma_{U}^2\tilde{\omega}^2]}>
+<\frac{\omega_{\rm p}^2[\tilde{\omega}_0 k_x U_y + U_x U_y/c^2 (\gamma_{U}^2 k_z^2 v_{\parallel}^2 
-\omega^2)]\gamma\gamma_{U}^2}
{\omega^2(\omega_B^2-\gamma^2\gamma_{U}^2\tilde{\omega}^2)}>,\\
\varepsilon_{xz} & = & - <\frac{\gamma_{U}^2 k_z U_{x}\omega^2_p(\omega-k_xU_{x})}
{\tilde{\omega}^2\gamma^3\omega^2}> 
+ <\frac{\omega_{\rm p}^2[(\tilde{\omega}_{0}-\omega U^2/c^2)v_{\parallel}(k_x - \omega U_{x}/c^2) 
+ k_z k_x v^2_{\parallel}U^2_y/c^2 ]\gamma\gamma_{U}^4}
{\omega^2(\omega_B^2-\gamma^2\gamma_{U}^2 \tilde{\omega}^2)}>  \nonumber \\
&& -i<\frac{\gamma_{U}^2 \omega_B\omega^2_p}
{\omega(\omega_B^2-\gamma^2\gamma_{U}^2\tilde{\omega}^2)}
\frac{U_x v_{\parallel}}{c^2}>,\\
\varepsilon_{yy} & = & 1-<\frac{k_z^2 \gamma_{U}^2 U^2_{y}\omega^2_p}{\tilde{\omega}^2\gamma^3\omega^2}>+<\frac{\omega_{\rm p}^2[\tilde{\omega}^2 + U_y^2/c^2 (k^2_z v^2_{\parallel} \gamma_{U}^2-\omega^2) + k^2_xU^2_y]\gamma\gamma_{U}^2}{\omega^2(\omega_B^2-\gamma^2\gamma_{U}^2\tilde{\omega}^2)}>,\\
\varepsilon_{yz} & = & - <\frac{k_z U_{y}\gamma_{U}^2\omega^2_p(\omega-k_xU_{x})}{\tilde{\omega}^2\gamma^3\omega^2}> 
+ <\frac{\omega_{\rm p}^2[k^2_x c^2- \omega^2 + \gamma_{U}^2 k_z v_{\parallel}(\omega - k_xU_x)]\gamma}{\omega^2(\omega_B^2 -\gamma^2\gamma_{U}^2\tilde{\omega}^2)}\frac{U_y v_{\parallel}\gamma_{U}^2}{c^2}> \nonumber \\
&& - i <\frac{\omega_B\omega^2_p\gamma_{U}^2}{\omega^2(\omega_B^2-\gamma^2\gamma_{U}^2\tilde{\omega}^2)}v_{
\parallel} (k_x - \frac{\omega U_x}{c^2})>,\\
\varepsilon_{zz} & = & 1 - <\frac{\omega_p^2(\omega-k_xU_{x})^2\gamma_{U}^2}{\omega^2\tilde{\omega}^2\gamma^3}> + <\frac{\omega_{\rm p}^2[(k_x c- \omega U_{x}/c)^2 + U_y^2/c^2(\omega^2-k^2_x c^2)]\gamma}{\omega^2(\omega_B^2 -\gamma_{U}^2\gamma^2\tilde{\omega}^2)}\frac{v_{\parallel}^2\gamma_{U}^4}{c^2}>, \\
{\rm Im}\,[\varepsilon_{yy}] & = & - \pi<\frac{\omega_{\rm p}^2[\tilde{\omega}^2+U_y^2/c^2(k^2_z v^2_{\parallel}\gamma_{U}^2-\omega^2)+ k^2_xU^2_y]\gamma_{U}}{\omega^2\tilde{\omega}}\delta(|\omega_B|-\gamma \gamma_{U}\tilde{\omega})>.
\end{eqnarray}
\end{widetext}
Here the quantity $\tilde{\omega}_{0} = \omega - k_z v_{\parallel}$. Other components of the
permittivity tensor are defined by its hermiticity. 
Therefore, in the expression for gyrofrequency
$$\omega_B=\frac{eB}{m_{\rm e} c}$$
we should take into the account the sign of the charge $e$. One can check that
moving to the appropriate reference frame this dielectric tesor reduces to the 
well-known one (Godfray et al. 1975; Suvorov \& Chugunov 1975; Hardee \& Rose 1975) 
obtained for the case ${\bf E} = 0$. Finally, as we see, the cyclotron resonance
corresponds to the condition $\omega_B = \gamma\gamma_{U}\tilde{\omega}$. 

As a result, one can find for the coefficients in Kravtsov-Orlov equations (\ref{t1})--(\ref{t2}):

\begin{widetext}
 
\begin{eqnarray}
\Lambda & = & - \frac{1}{2}<\frac{\omega^2_{\rm p}}{\tilde{\omega}^2\gamma^3}
\frac{\gamma_{U}^2\omega^2_B}{\omega_B^2-\gamma^2\gamma_{U}^2\tilde{\omega}^2}>
\left[(\sin \theta - U_x/c)^2 + \cos^2 \theta \, U^2_y/c^2\right],\\
{\rm Re}\,[\varepsilon_{x'y'}] & = & <\frac{\omega^2_{\rm p}}{\tilde{\omega}^2\gamma^3}
\frac{\gamma_{U}^2\omega^2_B}{\omega_B^2-\gamma^2\gamma_{U}^2\tilde{\omega}^2}>U_y/c
\left(\sin \theta - U_x/c\right)\cos \theta,\\
\varepsilon_{y'y'} - \varepsilon_{x'x'} & = &
<\frac{\omega^2_{\rm p}}{\tilde{\omega}^2\gamma^3}
\frac{\gamma_{U}^2\omega^2_B}{\omega_B^2-\gamma^2\gamma_{U}^2\tilde{\omega}^2}>
\left[(\sin \theta - U_x/c)^2 - \cos^2 \theta \, U^2_y/c^2\right],\\
{\rm Im}\,[\varepsilon_{x'y'}] & = & 
<\frac{\omega^2_{\rm p}}{\omega}
\frac{\gamma_{U}^2\omega_B}{\omega_B^2-\gamma^2\gamma_{U}^2\tilde{\omega}^2}>
\left[\cos \theta \gamma_{U}^{-2} 
- v_{\parallel}/c(1 - \sin \theta \, U_x/c )\right],\\
\tan (2\delta) & = & - \frac{2 U_y/c \, \cos \theta (\sin \theta - U_x/c) }
{(\sin \theta - U_x/c)^2 - \cos^2 \theta U^2_y/c^2},\\
\tan (\delta) & = & - \frac{\cos \theta \, U_y/c \, }{\sin \theta - {U_x}/c}.
\end{eqnarray}

\end{widetext}

\section{Rotation of coordinate systems}

The equations of the transition of the dielectric tensor from the coordinate system formulated
in Sect. 3 to that used in Sect. 4 are the following
\begin{eqnarray}
\varepsilon_{x'x'} & = & \varepsilon_{xx}\cos^2 \theta+\varepsilon_{zz}\sin^2 \theta 
- (\varepsilon_{xz}+\varepsilon_{zx})
\sin \theta  \, \cos \theta,
\nonumber \\
\varepsilon_{y'y'} & = & \varepsilon_{yy},
\nonumber \\
\varepsilon_{z'z'} & =& \varepsilon_{xx}\sin^2 \theta+\varepsilon_{zz}\cos^2 \theta 
+ (\varepsilon_{xz}+\varepsilon_{zx})
\sin \theta \, \cos \theta,
\nonumber \\
\varepsilon_{x'y'} & = & \varepsilon_{xy}\cos \theta -\varepsilon_{zy}\sin \theta,
\nonumber \\
\varepsilon_{y'z'} & = & \varepsilon_{yx}\sin \theta+\varepsilon_{yz}\cos \theta,
\nonumber \\
\varepsilon_{x'z'} & = & (\varepsilon_{xx}-\varepsilon_{zz})\sin \theta \,\cos \theta+\varepsilon_{xz}\cos^2 \theta-\varepsilon_{zx}\sin^2 \theta.
\nonumber 
\end{eqnarray}
Here $\theta$ is the angle between the ray propagation and local magnetic field directions. 
Other components of the tensor can be defined by its hermiticity.

\section{Natural modes}

For obtaining naturals modes of tensor $\varepsilon_{i'j'}$ it is convenient to rotate the 
coordinate frame in {\it xy}-plane to make the condition ${\rm Re}\,[\varepsilon_{x'y'}] = 0$ 
to be valid. One can 
find that it can be done by rotation defined by the following matrix:
\begin{equation}
\nonumber \\
M =
\nonumber \\
\pmatrix{
\cos a&& -\sin a && 0 \cr  
\sin a&& \cos a && 0 \cr
0&& 0 &&1},
\nonumber \\
\end{equation}
where 
\begin{equation}
\tan (2a) = \tan (2\delta) = - \frac{2{\rm Re}\,[\varepsilon_{x'y'}]}{\varepsilon_{y'y'}
-\varepsilon_{x'x'}}.
\end{equation}
In this frame the corresponding expressions for tensor components (defining as $\epsilon_{i'j'}$) 
are:
\begin{eqnarray}
\epsilon_{x'x'} & = & \varepsilon_{x'x'}\cos^2 a + \varepsilon_{y'y'}\sin^2 a 
+ 2 {\rm Re}\,[\varepsilon_{x'y'}]\sin a \, \cos a,
\nonumber \\
\epsilon_{y'y'} & = & \varepsilon_{x'x'}\sin^2 a + \varepsilon_{y'y'}\cos^2 a 
- 2 {\rm Re}\,[\varepsilon_{x'y'}]\sin a \,\cos a,
\nonumber \\
\epsilon_{x'y'} & = & i \, {\rm Im}\,[\varepsilon_{xy}].
\nonumber 
\end{eqnarray}
The naturals modes for this tensor (with pure imaginary {\it x'y'} component) are well-known 
(Andrianov \& Beskin 2010; Wang et al. 2010):
\begin{eqnarray}
\Theta_1 & = & \beta, \quad \quad \quad
{\rm sinh}2\Theta_2 = - \frac{1}{Q}, \\
\Theta_1 & = & \beta + \pi/2, \quad  
{\rm sinh}2\Theta_2 = \frac{1}{Q},
\end{eqnarray}
where by the definition
\begin{equation}
Q = i \, \frac{\epsilon_{y'y'} - \epsilon_{x'x'}}{2\epsilon_{x'y'}}.
\end{equation}
It can be easily shown that in our case 
\begin{equation}
\frac{1}{Q} = -\frac{{\rm Im}\,[\varepsilon_{x'y'}]}{\Lambda}.
\end{equation}
Hence, Eqns. (\ref{modes1})-(\ref{modes2}) really define the natural modes in the 
laboratory frame. The simple expression, which corresponds to a zero plasma temperature 
for $Q$ is the following (cf. Melrose \& Luo 2004 for zero drift):
\begin{equation}
Q = \frac{\lambda\omega_B\omega\left[(\sin \theta - U_x/c)^2 + U^2_y/c^2\cos^2 \theta\right]}
{2\gamma^3\tilde{\omega}^2\left[\cos \theta (1 - U^2/c^2) -{v_{\parallel}}/{c}(1 - \sin \theta \, {U_x}/{c})\right]}.
\end{equation}

\end{document}